\documentclass[a4paper,fleqn,usenatbib]{mnras}
\usepackage{newtxtext,newtxmath}
\usepackage{siunitx}
\usepackage[T1]{fontenc}
\usepackage{ae,aecompl}
\usepackage{graphicx}	
\usepackage{amsmath}	
\usepackage{amssymb}	
\usepackage{lscape}
\graphicspath{ {figures/} }

\title[Study of giant pulses from PSR\,B1937+21 using LEAP]{A detailed study of giant pulses from PSR\,B1937+21 using the Large European Array for Pulsars}

\author[McKee et al.]{J.\,W.\,McKee$^{1,2}$\thanks{E-mail: jmckee@mpifr-bonn.mpg.de},
B.\,W.\,Stappers$^{1}$,
C.\,G.\,Bassa$^{3}$,
S.\,Chen$^{4}$,
I.\,Cognard$^{5,6}$,
M.\,Gaikwad$^{2}$,
\newauthor
G.\,H.\,Janssen$^{3,7}$,
R.\,Karuppusamy$^{2}$,
M.\,Kramer$^{2,1}$,
K.\,J.\,Lee$^{8,2}$,
K.\,Liu$^{2,6}$,
D.\,Perrodin$^{9}$,
\newauthor
S.\,A.\,Sanidas$^{10,1}$,
R.\,Smits$^{3}$,
L.\,Wang$^{1,11}$,
and W.\,W.\,Zhu$^{11,2}$
\\
$^{1}$Jodrell Bank Centre for Astrophysics, School of Physics and Astronomy, The University of Manchester, Manchester M13 9PL, UK\\
$^{2}$Max-Planck-Institut f{\"u}r Radioastronomie, Auf dem H{\"u}gel 69, D-53121 Bonn, Germany\\
$^{3}$ASTRON, the Netherlands Institute for Radio Astronomy, Postbus 2, 7990 AA, Dwingeloo, The Netherlands\\
$^{4}$Institute of Gravitational Wave Astronomy and School of Physics and
Astronomy, University of Birmingham, Birmingham B15 2TT, UK\\
$^{5}$Laboratoire de Physique et Chimie de l'Environnement et de l'Espace LPC2E CNRS-Universit\'{e} d'Orl\'{e}ans, F-45071 Orl\'{e}ans, France \\
$^{6}$Station de Radioastronomie de Nan{\c c}ay, Observatoire de Paris, CNRS/INSU, F-18330 Nan{\c c}ay, France\\
$^{7}$Department of Astrophysics/IMAPP, Radboud University, P.O. Box 9010, 6500 GL Nijmegen, The Netherlands\\
$^{8}$Kavli institute for astronomy and astrophysics, Peking University, Beijing 100871, P.\,R.\,China\\
$^{9}$INAF - Osservatorio Astronomico di Cagliari, Via della Scienza 5, I-09047 Selargius (CA), Italy\\
$^{10}$Anton Pannekoek Institute for Astronomy, University of Amsterdam, Science Park 904, NL-1098 XH Amsterdam, The Netherlands\\
$^{11}$National Astronomical Observatories, Chinese Academy of Sciences, A20 Datun Rd, Chaoyang District, Beijing 100012, P.\,R.\,China
}

\date{Accepted XXX. Received YYY; in original form ZZZ}

\pubyear{2018}

\begin{document}
\label{firstpage}
\pagerange{\pageref{firstpage}--\pageref{lastpage}}
\maketitle

\begin{abstract}
We have studied 4265 giant pulses (GPs) from the millisecond pulsar B1937+21; the largest-ever sample gathered for this pulsar, in observations made with the Large European Array for Pulsars. The pulse energy distribution of GPs associated with the interpulse are well-described by a power law, with index $\alpha=-3.99\pm0.04$, while those associated with the main pulse are best-described by a broken power law, with the break occurring at $\sim7$\,Jy\,$\SI{}{\micro\second}$, with power law indices $\alpha_{\text{low}}=-3.48\pm0.04$ and $\alpha_{\text{high}}=-2.10\pm0.09$. 
The modulation indices of the GP emission are measured, which are found to vary by $\sim0.5$ at pulse phases close to the centre of the GP phase distributions. We find the frequency-resolved structure of GPs to vary significantly, and in a manner that cannot be attributed to the interstellar medium influence on the observed pulses. 
We examine the distribution of polarisation fractions of the GPs and find no correlation between GP emission phase and fractional polarisation. We use the GPs to time PSR\,B1937+21 and although the achievable time of arrival precision of the GPs is approximately a factor of two greater than that of the average pulse profile, there is a negligible difference in the precision of the overall timing solution when using the GPs.
\end{abstract}



\section{Introduction}
\begin{figure}
	\includegraphics[scale=0.175]{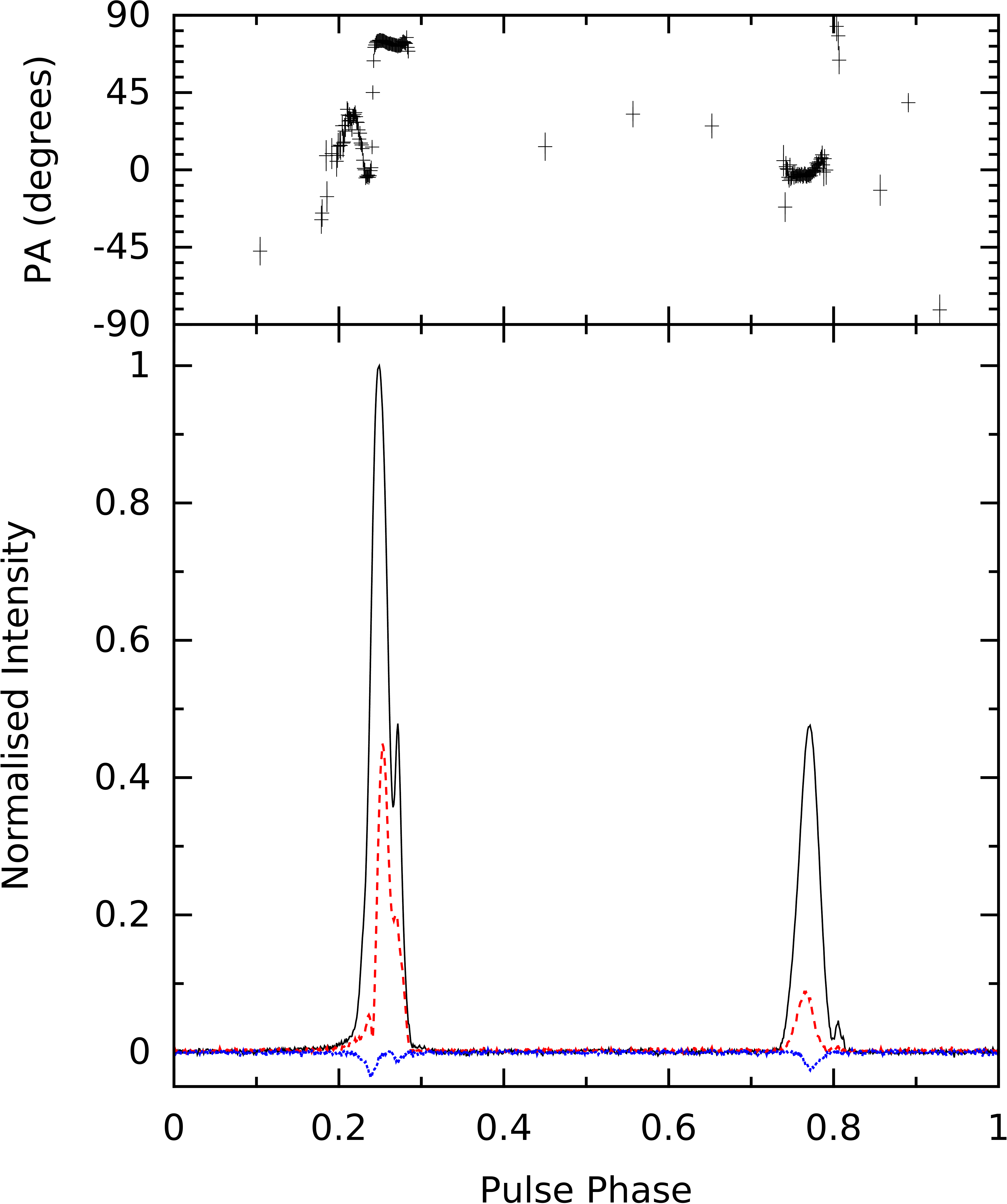}	
	\centering
	\caption{PSR\,B1937+21 coherently-added LEAP profile from observations made on 22$^{\text{nd}}$ May 2014 (MJD\,56799), with the polarisation position angles displayed in the top panel. The higher- and lower-amplitude peaks are referred to as the main pulse (MP) and interpulse (IP) respectively. The line types indicate the Stokes parameters: total intensity (I, solid black), total linear (L, dashed red), and total circular (V, dotted blue).}
	\label{fig:B1937polprof}
\end{figure}
\begin{figure}
	\includegraphics[scale=0.35]{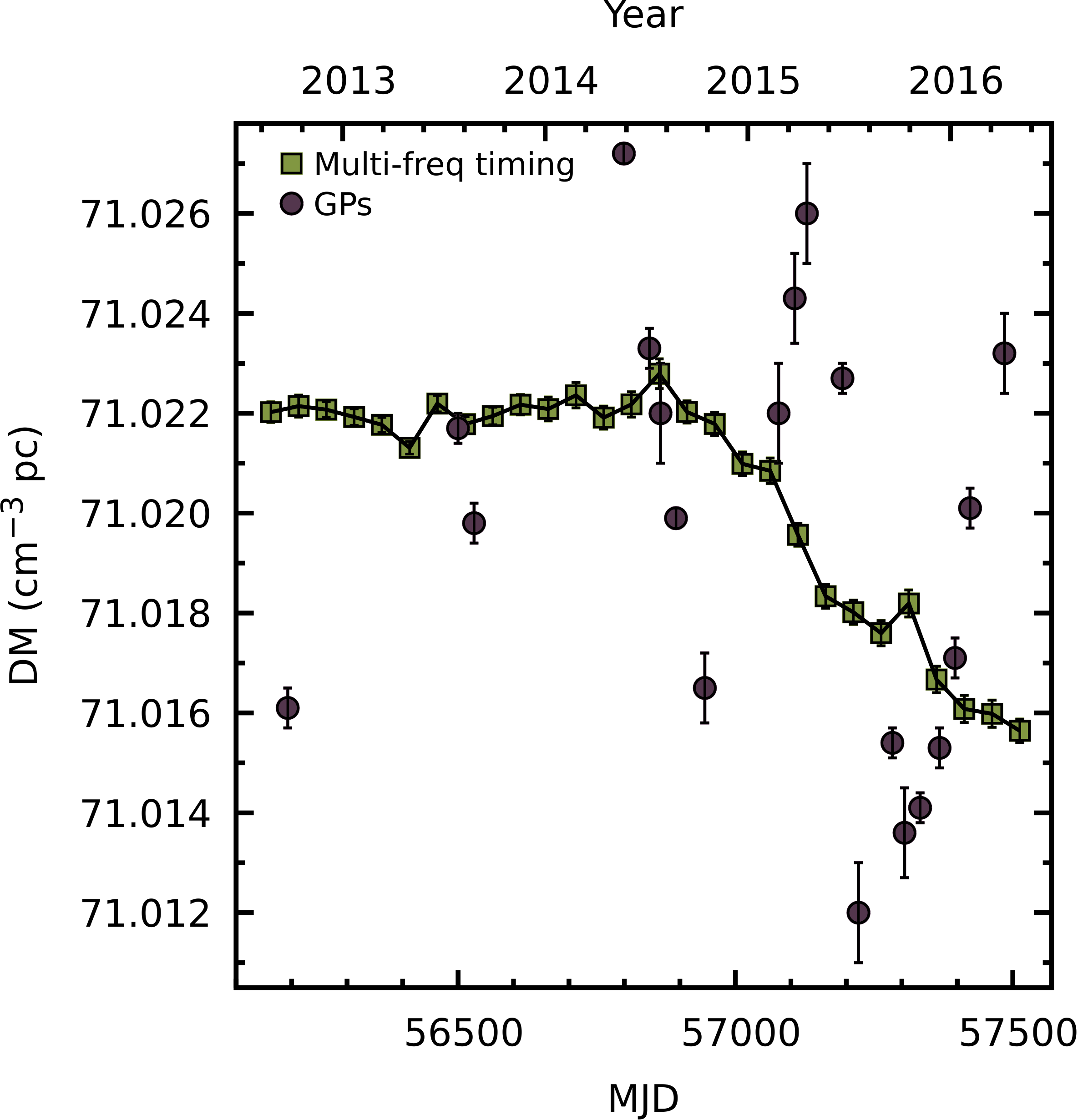}
	\centering
	\caption{Measured DMs for PSR\,B1937+21 from multi-frequency timing (yellow squares, dashed line) used as the initial guess for the value used in the dedispersion process, and the mean DM measured from GPs from each observation (purple circles). The initial DMs were derived by fitting for DM over 50-day windows in a data set composed of TOAs from the Lovell Telescope's ROACH backend at a centre frequency of 1532\,MHz with 400\,MHz of bandwidth, and from the Jodrell Bank 42-ft telescope's COBRA2 backend at a centre frequency of 610\,MHz, with 10\,MHz of bandwidth. The 42-ft telescope data are obtained almost daily, and the Lovell Telescope data approximately every 1-2 weeks.}
	\label{fig:B1937multifreqdm}
\end{figure}

PSR\,B1937+21 (PSR\,J1939+2134) was the first millisecond pulsar (MSP) to be discovered \protect\citep{bkh+82}, and with a period of 1.56\,ms, was the fastest-spinning known pulsar, until the discovery of the 1.40\,ms PSR\,J1748$-$2446ad \protect\citep{hrs+06}. After the Crab Pulsar (PSR\,B0531+21, \protect\citealp{sr68}, \protect\citealp{hc70}, \protect\citealp{ss70}), PSR\ B1937+21 was the second pulsar found to exhibit so-called `giant pulse' (GP) emission (\protect\citealp{wcs84}, \protect\citealp{cst+96}); occasional very short-duration single pulses with flux densities greatly exceeding that of the single-pulse average. GPs are thought to be generated by a different emission mechanism to that of the regular pulsed emission, and originate from a different magnetospheric height, due to the following observational evidence:
\begin{enumerate}
\item The pulse width of individual GPs is typically much narrower than that of regular single pulses, with (unresolved) durations as short as 2\,ns, in the case of the Crab Pulsar \protect\citep{hkw+03}. 
\item The pulse energy distributions of the regular emission are usually well-modelled as a log-normal distribution, while pulse energies of GPs are found to follow a power law distribution (\protect\citealp{ag72}, and see more recent examples by \protect\citealp{cbh+04}, \protect\citealp{oob+15}).
\item GP emission often occurs in a narrow phase window \protect\citep{kt00}, which in most cases is offset from the regular pulse emission region, and is often found to be phase-aligned with high-energy (X-ray and $\gamma$-ray) emission \protect\citep{rj01}, which suggests that GPs may be a radio component of the high-energy emission (\citealp{chk+03}, \citealp{sso+03}). 
\end{enumerate}

PSR\,B1937+21 is considered to be an orthogonal rotator \protect\citep{stc99}, with two main emission regions separated by $\sim0.5$ in pulse phase (Figure \ref{fig:B1937polprof}). The pulsar has GP emission regions associated with the trailing edge of both the main pulse (MP) and interpulse (IP) emission regions, which are coincident with the relative pulse phase of the X-ray emission in both pulse profile components \protect\citep{chk+03}. To differentiate between these two emission regions, we use the terms MGP and IGP to refer to GPs from the MP and IP respectively, and use GP to refer to the phenomenon regardless of pulse phase.

\begin{table*}
\caption{Summary of the 21 LEAP observations of PSR\,B1937+21 used in this work, and the number of GPs from each observation. Note that two separate observations were made on 2012-09-23, separated by $\sim13$\,mins. S/N$_{\text{profile}}$ refers to the integrated signal-to-noise of the average pulse profile. The `Telescopes' heading refers to the telescopes included in the combined LEAP observation. \textit{Telescope code:} E: Effelsberg Telescope; J: Jodrell Bank (Lovell Telescope); N: Nan\c{c}ay Radio Telescope; S: Sardinia Radio Telescope; W: Westerbork Synthesis Radio Telescope.}
\label{tab:b1937observations}
\centering 
\begin{tabular} {c c c c c c c c c c c}
\hline 
Date & MJD & Telescopes & $f_{\text{c}}$ (MHz) & BW (MHz) & $T_{\text{obs}}$ (sec) & S/N$_{\text{profile}}$ & Coherency & $N_{\text{MGP}}$ & $N_{\text{IGP}}$ & $N_{\text{GP}}$ \\
\hline
\hline
2012-09-23 (1) & 56193 & EJW & 1412 & 64 & 1240& 719 & 80$\%$ & 125 & 77 & 202 \\
2012-09-23 (2) & 56193 & EJW & 1412 & 64 &1020 & 808 & 86$\%$ & 124 & 74 & 198 \\
2013-07-27 & 56500 & EJW & 1404 & 112 & 1800 & 525 & 63$\%$ & 138 & 106 & 244 \\
2013-08-25 & 56529 & EW & 1364 & 64 & 2700 & 1065 & 98$\%$ & 278 & 154 & 432 \\
2014-05-22 & 56799 & ESW & 1396 & 128 & 2760 & 904 & 86$\%$ & 168 & 83 & 251 \\
2014-07-07 & 56845 & EW & 1420 & 80 & 2770 & 435 & 88$\%$ & 172 & 84 & 256 \\
2014-07-27 & 56865 & EW & 1380 & 96 & 2760 & 492 & 97$\%$ & 31 & 22 & 53 \\
2014-08-24 & 56893 & EJNW & 1396 & 128 & 2490 & 852 & 92$\%$ & 155 & 75 & 230 \\
2014-10-15 & 56945 & EJNW & 1380 & 64 & 1470 & 587 & 57$\%$ & 139 & 76 & 215 \\
2015-02-25 & 57078 & EJNW & 1396 & 128 & 2120 & 715 & 76$\%$ & 145 & 94 & 239 \\
2015-03-26 & 57107 & EJNSW & 1396 & 128 & 2510 & 1232 & 82$\%$ & 171 & 87 & 258 \\
2015-04-17 & 57129 & EJNSW & 1396 & 128 & 2510 & 1202 & 80$\%$ & 137 & 52 & 189 \\
2015-06-20 & 57193 & EJW & 1396 & 128 & 2230 & 598 & 100$\%$ & 112 & 63 & 175 \\
2015-07-19 & 57222 & EJS & 1412 & 96 & 2790 & 458 & 67$\%$ & 112 & 92 & 204 \\
2015-09-18 & 57283 & EJN & 1404 & 112 & 2520 & 429 & 78$\%$ & 90 & 32 & 122 \\
2015-10-10 & 57305 & EJ & 1412 & 96 & 2790 & 543 & 86$\%$ & 112 & 52 & 164 \\
2015-11-07 & 57333 & EJS & 1396 & 128 & 2790 & 729 & 99$\%$ & 81 & 55 & 136 \\
2015-12-12 & 57368 & EJNS & 1396 & 128 & 2520 & 755 & 75$\%$ & 138 & 55 & 193 \\
2016-01-09 & 57396 & EJNS & 1396 & 128 & 2520 & 1122 & 83$\%$ & 154 & 99 & 253 \\
2016-02-05 & 57423 & EJNS & 1404 & 112 & 2390 & 629 & 75$\%$ & 97 & 49 & 146 \\
2016-04-07 & 57485 & EJNS & 1388 & 80 & 2380 & 924 & 78$\%$ & 110 & 48 & 158 \\
\hline
Mean & - & - & - & - & 2337 & 709 & 82$\%$ & 133 & 70 & 203 \\
\hline
Total & - & - & - & - & 49080 & - & - & 2789 & 1476 & 4265 \\
\hline
\end{tabular}
\end{table*}

PSR\,B1937+21 is one of the most luminous MSPs discovered to date, and is included in all currently-ongoing pulsar timing arrays (\protect\citealp{dcl+16}, \protect\citealp{abb+18}, \protect\citealp{rhc+16}, \protect\citealp{vlh+16}). It is one of the few MSPs that is presently known to exhibit significant timing noise (\protect\citealp{ktr94}, \protect\citealp{scm+13}, \protect\citealp{cll+16}), in addition to large variations in dispersion measure (DM) and scattering (\protect\citealp{rdb+06}, \protect\citealp{kcs+13}, \protect\citealp{lmj+16}), all of which impose limitations on the achievable timing precision of the pulsar. It has been shown that the non-GP single-pulse emission from PSR\,B1937+21 is extremely regular, with no evidence for intrinsic pulse-shape variations over millions of rotations \protect\citep{jap01}, 
and information gathered from individual rotations of the pulsar may shed light on the nature of the pulsar's timing noise. 

The structure of our paper is as follows: in Section \ref{observations_section}, we describe our observations, data reduction, and calibration process. In Section \ref{GP_pipeline}, we describe the method we have used to search for GPs in our data set. In Section \ref{results_section}, we present the results of our GP search, and show results of analyses using our GP data set. We discuss our findings in Section \ref{discussion_section}, and make general concluding comments in Section \ref{conclusions_section}.

\section{Observations} \label{observations_section}
Our observations were made using the Large European Array for Pulsars (LEAP); a tied-array telescope comprised of various combinations of five European 100-m class radio telescopes: the Effelsberg Telescope, the Lovell Telescope, the Nan\c{c}ay Radio Telescope (NRT), the Westerbork Synthesis Radio Telescope (WSRT), and the Sardinia Radio Telescope (SRT). Observations with the participating telescopes are made simultaneously and are combined coherently into a tied-array beam, forming a virtual 195-m dish when using all five telescopes. 
The design of the LEAP experiment is detailed in \protect\cite{bjk+16}.

\begin{figure}
	\includegraphics[scale=0.355]{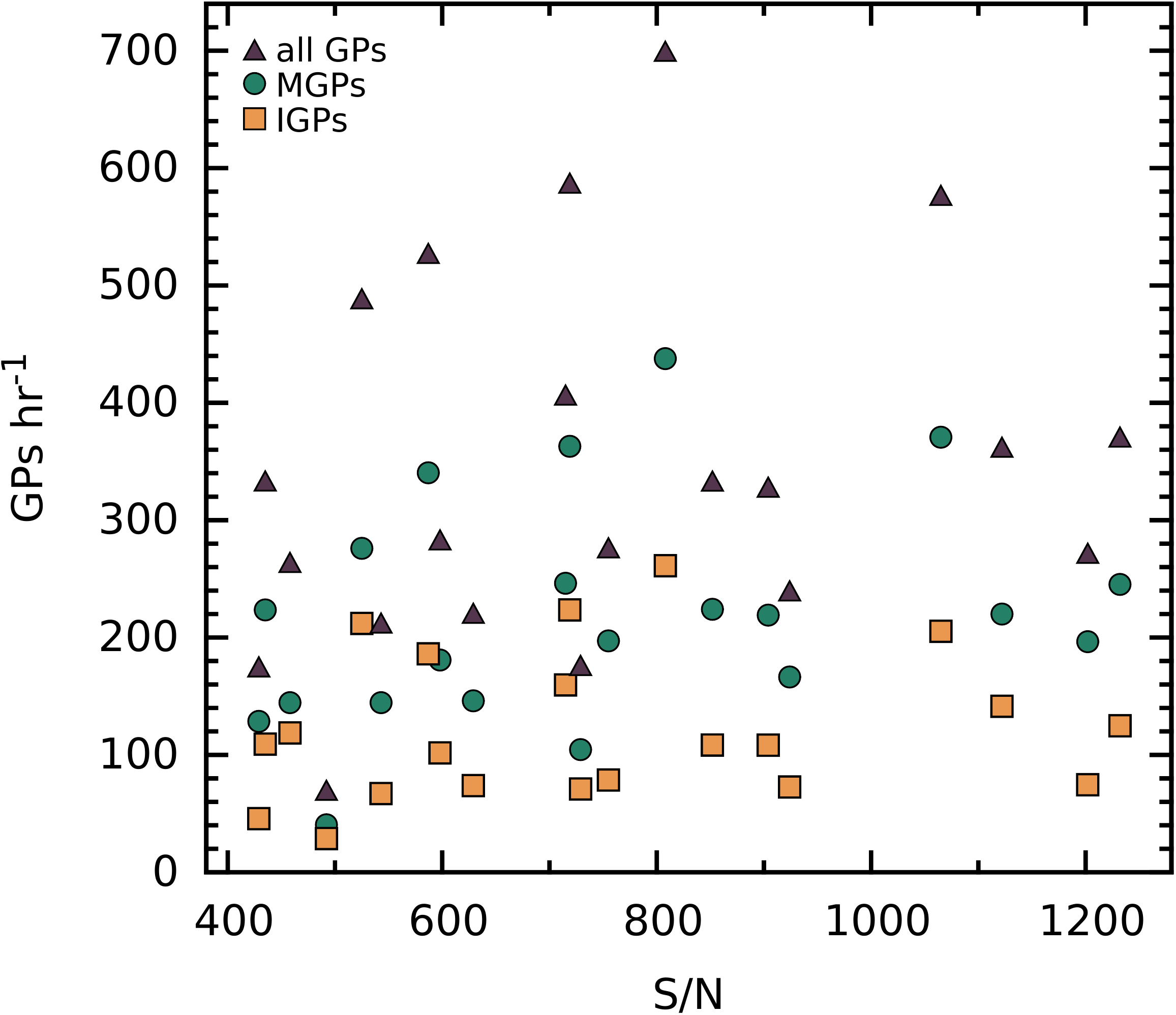}
	\centering
	\caption{Number of GPs vs. integrated profile S/N for the corresponding observation. There is no consistent correlation between the two quantities, indicating that the number of GPs is not S/N-limited. IGP: orange squares, MGP: green circles, total GPs: purple triangles.}
	\label{fig:GPsnrvsnumber}
\end{figure}
\begin{figure*}
	\includegraphics[scale=0.282]{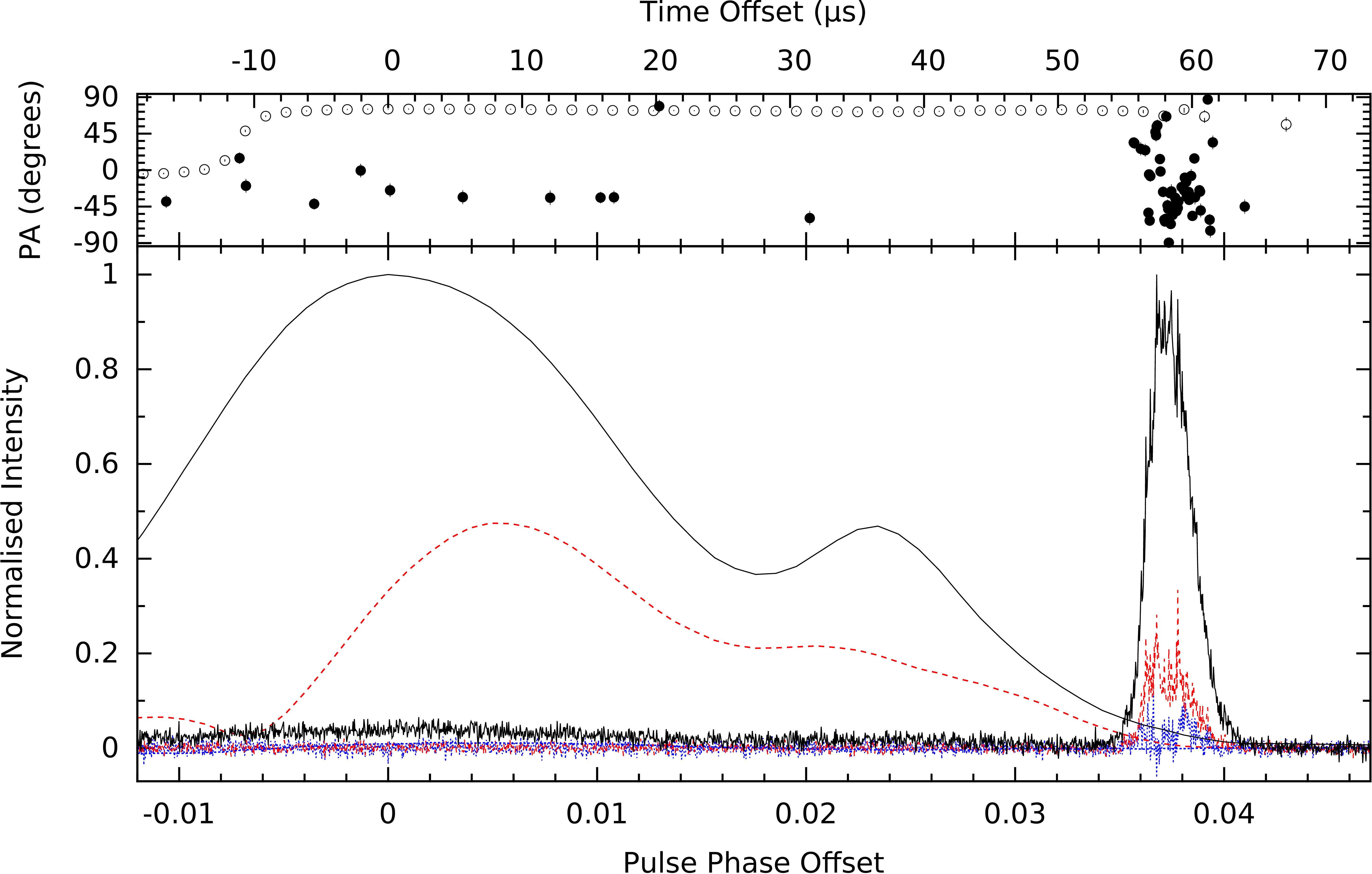} \\
          \vspace{0.3cm}
	\includegraphics[scale=0.282]{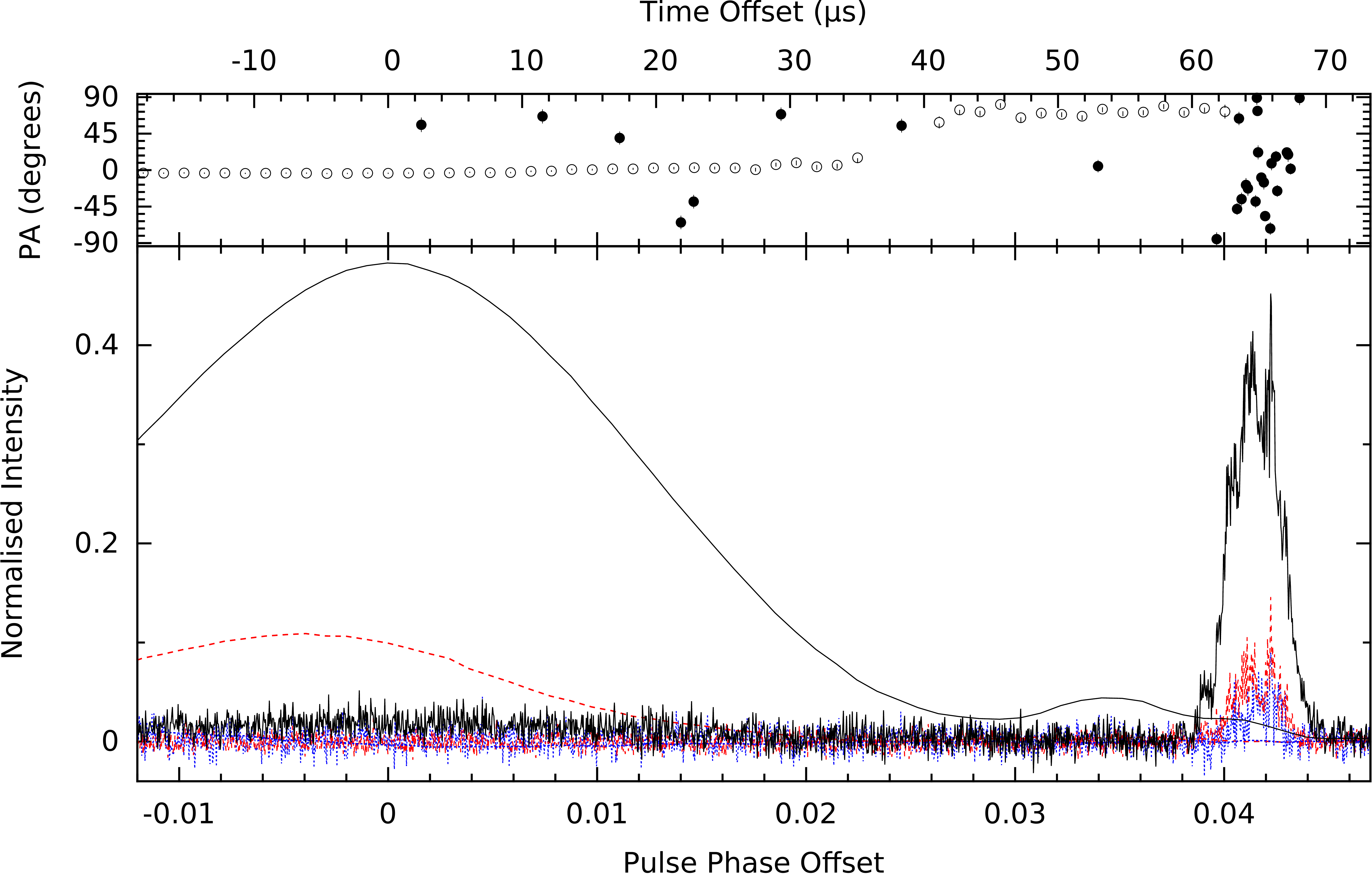}
	\centering
	\caption{Pulse intensity as a function of pulse phase of the regular emission using 1024 phase bins (centred at 0) and the GP emission using 8192 phase bins, averaged over all 21 observations, with the polarisation position angles displayed in the top panels for the regular emission (open circles) and the GP emission (closed circles). The profiles are normalised so that the peak MP and MGP fluxes are unity, and the profiles are rotated so that the peaks of the MP and IP are centred at zero phase. The peak of the MGP trails the peak of the MP by approximately \SI{58}{\micro\second} (above) and the peak of the IGP trails the peak of the IP by approximately \SI{64}{\micro\second} (below). In both plots, a small contribution from the normal emission is visible in the GP data, indicating that normal emission occurs simultaneously with GP emission. The line types indicate the Stokes parameters: total intensity (I, solid, black), total linear (L, dashed, red), and total circular (V, dotted, blue).}
	\label{fig:averageprofile}
\end{figure*}
\begin{figure}
	\includegraphics[scale=0.355]{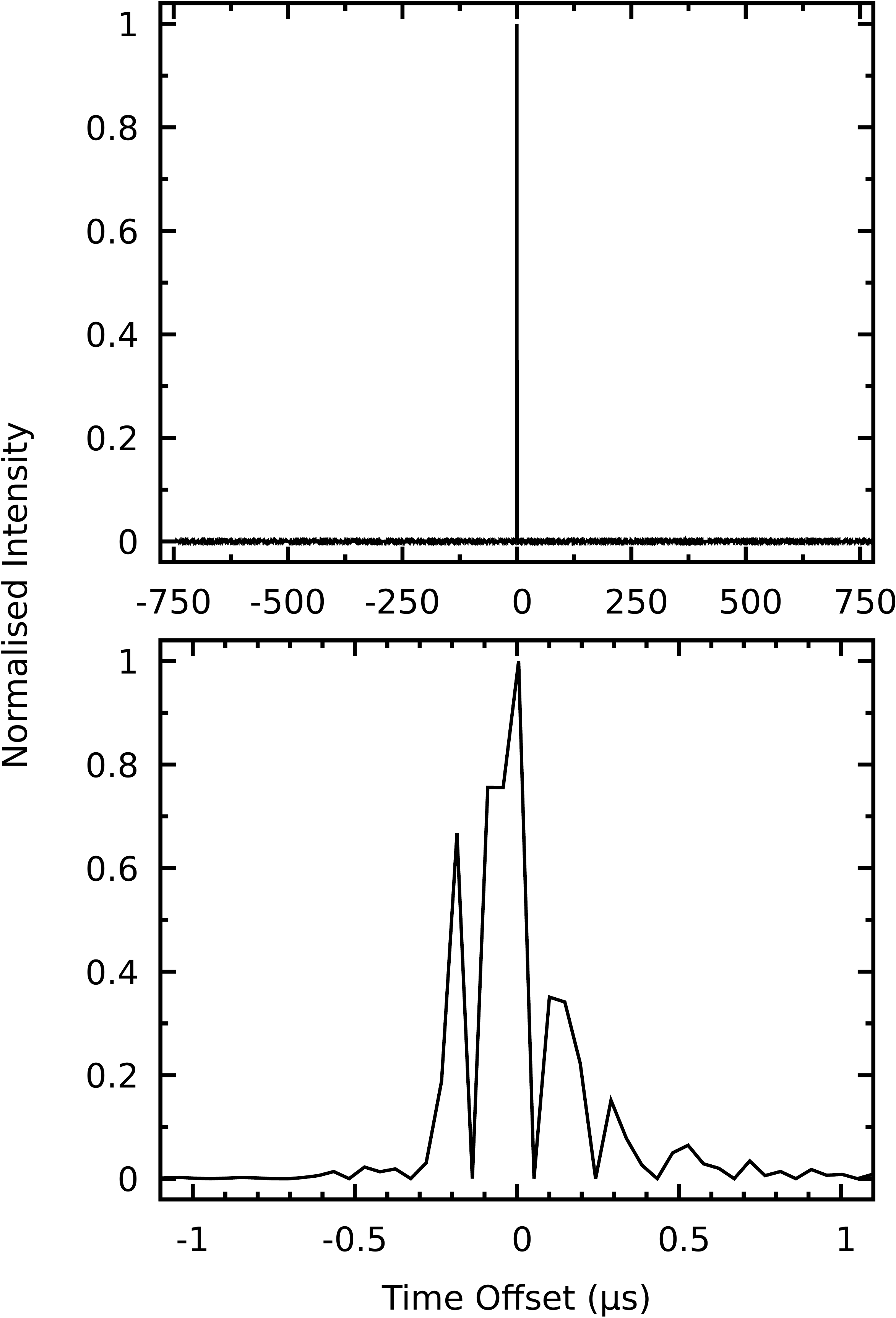}	
	\centering
	\caption{Total-intensity profile of the brightest GP observed in our search. The observation was made on 22$^{\text{nd}}$ May 2014 (MJD\,56799), at a time resolution of $\sim47$\,ns. The integrated signal-to-noise ratio of the pulse is 1988, and the pulse energy is $492$\,Jy\,$\SI{}{\micro\second}$. The upper plot shows the full pulse phase, and the lower plot is zoomed in to 0.0014 of the pulse phase (\SI{2.2}{\micro\second}). The effect of scattering on this pulse is very small, with a measured scattering time scale $\tau_{\text{sc}}=128\pm8$\,ns and a pulse width at $1\%$ of the maximum of $\sim\SI{1.5}{\micro\second}$}. 
	\label{fig:brightGP}
\end{figure}
\begin{figure*}
	\includegraphics[scale=0.365]{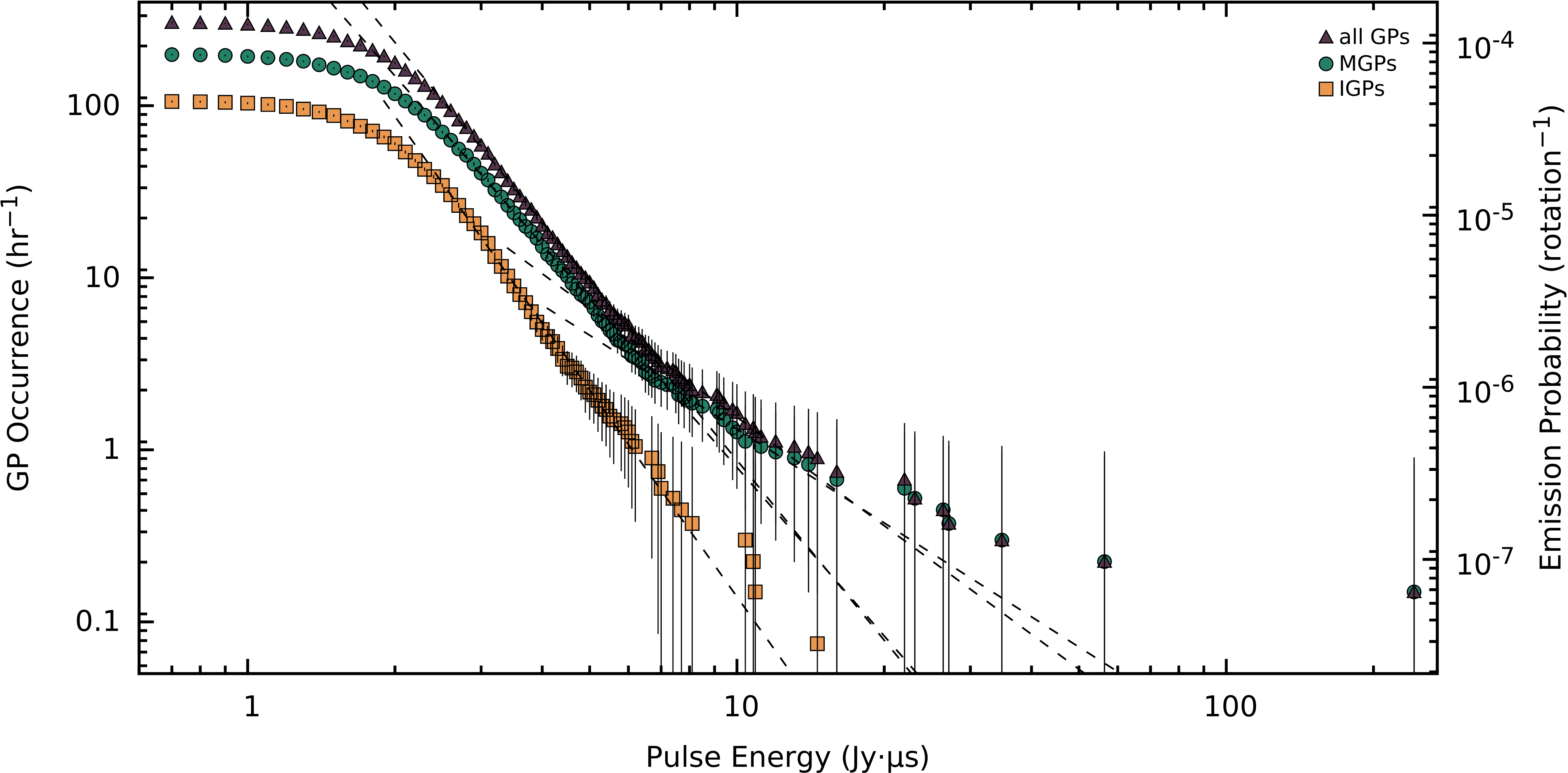}
	\centering
	\caption{Cumulative occurrence rates and emission probabilities for GPs exceeding a given pulse energy, binned in intervals of 0.1\,Jy\,$\SI{}{\micro\second}$. The pulse energies of the MGPs (green circles), IGPs (orange squares), and all GPs (purple triangles) are plotted separately, with counting errors assigned as the standard error for a Poisson process. The MGP and all-GP distributions clearly exhibit a broken power law for GPs exceeding pulse energies of $\sim7$\,Jy\,$\SI{}{\micro\second}$. A broken power law is not visible in the IGP distribution, possibly due to the lower number of IGPs exceeding this pulse energy. The dashed lines are the best-fit power laws for pulse energies greater than the flattened low-energy regime, whose indices are listed in Table \ref{tab:powerlawindices}.}
	\label{fig:gp_flux_numbers}
\end{figure*}
Monthly observations have been made in LEAP mode since February 2012, and a summary of the observations analysed here is presented in Table \ref{tab:b1937observations}.
This includes details of the telescopes included in the observing sessions, the central frequency (typically 1396\,MHz) and bandwidth (typically 128\,MHz), and the observing duration. Differences in bandwidth are due to either the instrumental setup used in the observation or due to data loss, and in all cases the bandwidth uses contiguous sub-bands. 
Following an observation, the baseband data (raw complex voltages) from the individual telescopes are shipped to Jodrell Bank Observatory (in the case of NRT, data are transferred electronically), where the signals are correlated on a CPU cluster, using a software pipeline detailed in \protect\cite{sbj+17}. By correlating the single-telescope data, initially on a nearby calibrator and later on the pulsar itself, time and phase delays are measured between pairs of telescopes, and used to coherently add the single-telescope data to form the LEAP tied-array beam. We define the coherency of a correlated observation as the ratio of the S/N of the combined LEAP profile to the sum of the individual telescope S/Ns. An ideal addition will have a coherency of $100\%$, while an incoherent addition (i.e. simply adding the signals without using the phase information of the raw voltages) will have a coherency of $\sim N_{\text{tels}}^{-1/2}$, in cases where the instruments and noise are identical.
Prior to September 2014, data recorded during LEAP observations were archived for future reduction in monthly correlation campaigns. Currently, data are reduced at most two months after an observation, while archived observations are analysed in parallel. As the reduction of the archived data is still ongoing, there are gaps in our reduced data sets prior to May 2014.

For each observation, the data were polarisation-calibrated using baseband data from observations of PSR\,J1022+1001 and/or PSR\,B1933+16, made during the same LEAP observing session, with corresponding polarisation templates obtained from the European Pulsar Network database\footnote{\url{www.epta.eu.org/epndb}} \protect\citep{stc99}. An example of a polarisation-calibrated coherently-added pulse profile is shown in Figure \ref{fig:B1937polprof}.
Our polarisation calibration procedure uses a template matching algorithm, similar to the method developed by \protect\cite{van06}, which we discuss in detail in \protect\cite{bjk+16}. Coherently-added baseband data for the combined LEAP observations are saved to tape for future scientific use.

\section{Giant Pulse Search Pipeline} \label{GP_pipeline}
Our 21 observations recorded a total of $3.1\times10^{7}$ rotations of PSR\,B1937+21 (equivalent to $13.6$\,hrs of observations), and each rotation was searched for GPs. 
The LEAP baseband data are split into 10-second sub-integrations for each 16-MHz sub-band. These files were concatenated in time and frequency to obtain a single file per observation.
During the search, the frequency channels were summed, and both polarisations were summed in quadrature in each rotation. The DM used for coherent dedispersion was chosen using an initial value derived from multi-frequency observations of PSR\,B1937+21 using the Lovell Telescope at 1532\,MHz and the Jodrell Bank 42-ft telescope at 610\,MHz (Figure \ref{fig:B1937multifreqdm}). This value was then used to dedisperse data from the first few minutes of the observation, while high S/N ($>15\sigma$) GP candidates were searched for (full details of the search follow this paragraph). These initial candidates were then used to re-calibrate the DM by finding the values that minimised the pulse width of each of these GPs, and taking the final DM as the mean, and the RMS of the values as the uncertainty. Errors in the DM used to dedisperse the observation can smear out the signal, and lead to a decreased sensitivity to GPs. However, in the case where the full 128\,MHz bandwidth is used, we calculate a maximum smearing of $\sim0.4\,\SI{}{\micro\second}$ for a DM error of 0.001\,cm$^{-3}$\,pc (i.e. similar to the maximum uncertainty of our DM values used for folding), which does not significantly impact our sensitivity. 
We note that although we have used the same DM value to dedisperse both the MGPs and IGPs, the true quantity may not be consistent between GPs from each component, which potentially include different contributions from the magnetosphere \protect\citep{he07}.
The full search was then performed on the entire observation, which was coherently dedispersed with the optimised DM  using \textsc{DSPSR}\footnote{\url{http://dspsr.sourceforge.net/}} \protect\citep{vb11}.

Our DM values obtained from minimising the pulse width of bright GPs (described above) differ from those obtained from the multi-frequency timing (Figure \ref{fig:B1937multifreqdm}). This may indicate that the DM value that best-optimised our GP search over a small 64 to 128\,MHz bandwidth does not optimally describe the DM for observations at widely-separated observing frequencies, or that there is frequency-dependence of the GP structure that contaminates the DM measurement. Scattering and scintillation were not accounted for when choosing the optimal DM. These effects can also cause a frequency-dependent broadening of the pulse and therefore contaminate the DM measurement, but the magnitude is not sufficient to significantly alter measured DMs at our observing frequencies and over our small bandwidth. The DM values obtained via both techniques are in agreement with that of \protect\cite{ps03} (71.025\,cm$^{-3}$\,pc), and the portion of the data set which overlaps with measurements made using the European Pulsar Timing Array data release 1.0 ($\sim71.02$\,cm$^{-3}$\,pc, \protect\citealp{dcl+16}), but are consistently lower than those of \protect\cite{cbl+95} (71.041\,cm$^{-3}$\,pc) and \protect\cite{spb+04} ($71.036\pm0.004$\,cm$^{-3}$\,pc), which is expected, given the large known DM variations in this pulsar.

The entire data set was divided into single pulses, each with 8192 phase bins and a time resolution of $\sim47$\,ns. This slightly oversampled the data, which has an intrinsic resolution of 62.5\,ns, based on the 16\,MHz-wide bands. 
As scattering reduces this peak S/N of pulses, each rotation was searched for pulses that exceeded a S/N threshold of $7\sigma$ at three time resolutions of approximately 47\,ns, 95\,ns, and 190\,ns, corresponding to the initial resolution, and downsampling by factors of two and four respectively to ensure that scattered GPs were not missed. This also ensured that the diminished signal at high time resolutions did not cause GPs with low flux densities or sub-pulse structure to be missed.
Candidates were further selected by generating an intensity vs. phase plot from each rotation, and excluding rotations with a peak intensity which occurred outside of a pulse phase window of width 0.06, centred on the peak of the intensity vs. phase distribution. 
This final set of candidates was then inspected by eye, using total-intensity vs. pulse phase and frequency channel vs. pulse phase plots, to ensure that the signal was not narrow-band RFI (i.e. located in a single band), and that false positives from time-varying broadband RFI were removed. 

\section{Results} \label{results_section}
\subsection{Giant Pulses} \label{GPpulsesection}
Our search yielded 2789 MGPs and 1476 IGPs, for a total of 4265 GPs, which is the largest sample of GPs ever gathered for PSR\,B1937+21. Our sample has a mean MGP:IGP ratio of 65:35, which is consistent with the findings of \protect\cite{spb+04}, who observed an MGP:IGP ratio of 61:39 for a GP sample size of 309, from a 39-min observation in May 1999. The mean GP occurrence rate across all of our observations was 205\,hr$^{-1}$ for MGPs, and 108\,hr$^{-1}$ for IGPs, for a total of 313\,hr$^{-1}$. This is lower than the rate of 475\,hr$^{-1}$ found by \protect\cite{spb+04}, although we note that there appears to be intrinsic variability in the emission rate, with the rate varying by approximately an order of magnitude across all our observations (from 69\,hr$^{-1}$ to 699\,hr$^{-1}$). 

We observe no correlation between integrated pulse profile S/N and number of observed GPs (Figure \ref{fig:GPsnrvsnumber}), which indicates that the variation in emission rate is intrinsic, rather than related to diffractive interstellar scintillation, and that our detection rate is very close to the emission rate at our sensitivities. The number of GPs from each observation is summarised in Table \ref{tab:b1937observations}. As expected following \protect\cite{kt00}, the GP emission was found to occur in narrow phase windows (much narrower than those used for our candidate selection) located at the trailing edges of the regular emission regions, with the MGP trailing the MP peak by $\sim\SI{58}{\micro\second}$, and the IGP trailing the IP peak by $\sim\SI{64}{\micro\second}$ (Figure \ref{fig:averageprofile}). 

We note that emission at the phase of the regular average pulse profile is clearly visible in the total GP profile (Figure \ref{fig:averageprofile}), which demonstrates that regular emission occurs simultaneously with GP emission. 
As the GP emission regions are located at the edge of the regular emission regions, and a `bump' in the regular IP emission is coincident with the location of the IGP emission phase, the GP contribution to the regular average pulse profile was investigated. 
After scaling the total GP profile relative to the average pulse profile by a factor $(N_{\text{rotations}}-N_{\text{GP}})^{-1/2}$, it was found that the IGP contribution to the flux ratio at the phase of the IP `bump' is $\sim0.03$, and therefore negligible. 
Further investigation will be required to determine whether 
there is a long tail of `weak' GPs that are contributing to the flux at this phase, or if the `bump' is simply a feature of the underlying non-GP emission at this phase. 
We observe that in the case of high-S/N GPs, resolvable low-intensity emission precedes the leading edge of the main peaks (Figure \ref{fig:brightGP}), which indicates that GP emission in PSR\,B1937+21 does not occur in a single short-duration burst, but in a manner similar to the `nano-shots' observed in GP emission from the Crab Pulsar (\protect\citealp{hkw+03}, \protect\citealp{he07}).
\begin{table}
\large
\caption{Power law indices ($\alpha$) for the pulse energy distributions presented in Figure \ref{fig:gp_flux_numbers}. The power law regime is obeyed for pulse energies $\gtrsim2$\,Jy\,$\SI{}{\micro\second}$, below which the distribution is flattened. The IGP distribution is well-described by a single power law, while the MGP and all-GP distributions are better-described by a broken power law, with indices listed here as `low' and `high'.}
\label{tab:powerlawindices}
\centering 
\begin{tabular} {l l c}
\hline 
Data set & Pulse energy (Jy\,$\SI{}{\micro\second}$) & $\alpha$ \\
\hline
\hline
IGP & $\gtrsim2$ & $-3.99\pm0.04$ \\
\hline
MGP$_{\text{low}}$ & $\sim2$--$7$ & $-3.26\pm0.03$ \\
MGP$_{\text{high}}$ & $\gtrsim7$ & $-1.81\pm0.06$ \\
\hline
GP$_{\text{low}}$ & $\sim2$--$7$ & $-3.48\pm0.04$ \\
GP$_{\text{high}}$ & $\gtrsim7$ & $-2.10\pm0.09$ \\
\hline
\end{tabular}
\end{table}
\begin{figure*}
	\includegraphics[scale=0.255]{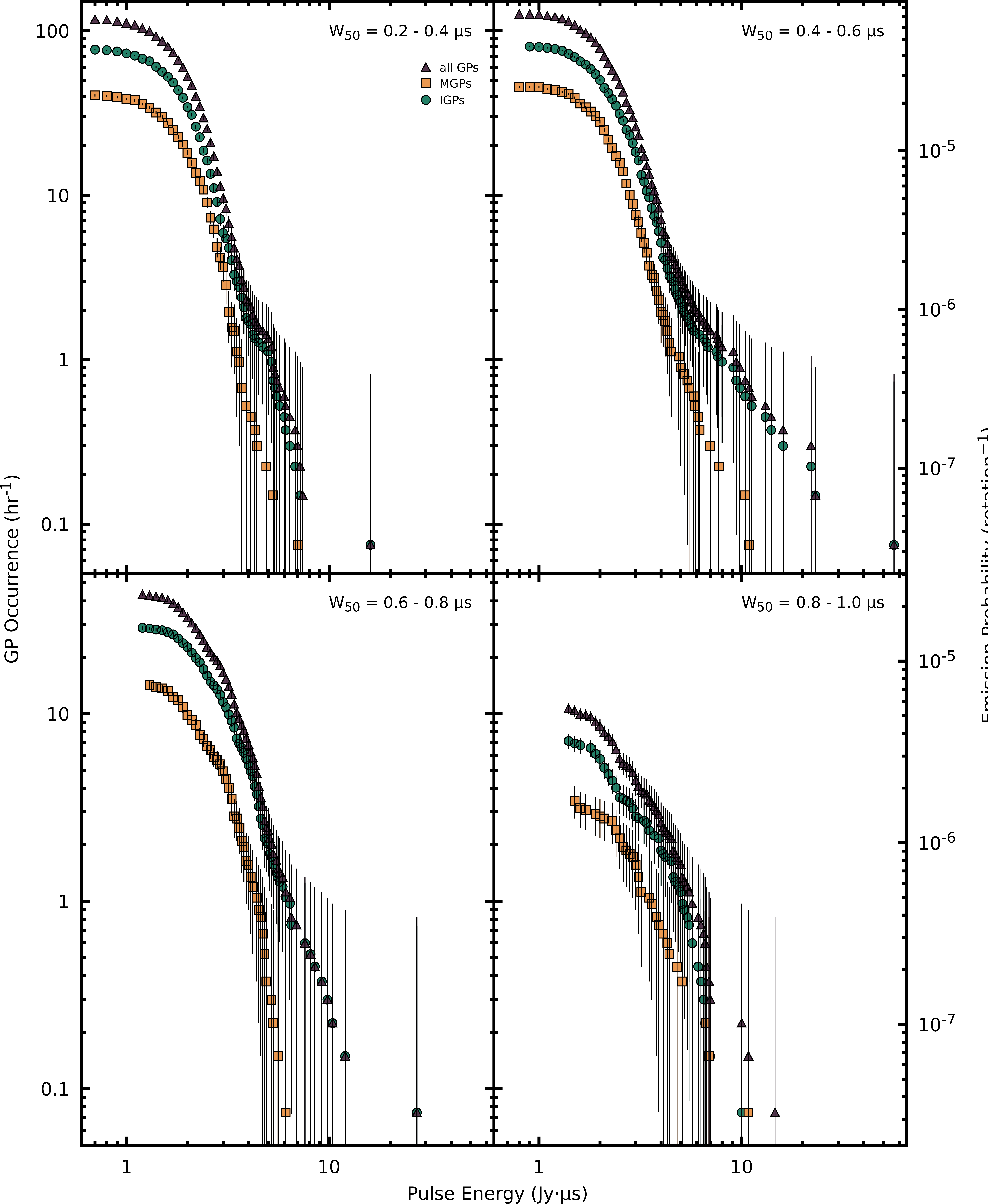}
	\centering
	\caption{As Figure \ref{fig:gp_flux_numbers}, for GP pulse energies at a given full-width at half-maximum ($W_{50}$). The break in the power law occurs at pulse energies of $\gtrsim7$\,Jy\,$\SI{}{\micro\second}$ for the $0.2\,\SI{}{\micro\second}<W_{50}<0.4\,\SI{}{\micro\second}$, $0.4\,\SI{}{\micro\second}<W_{50}<0.6\,\SI{}{\micro\second}$ and $0.6\,\SI{}{\micro\second}<W_{50}<0.8\,\SI{}{\micro\second}$ distributions, and is not easily distinguishable in the $0.8\,\SI{}{\micro\second}<W_{50}<1.0\,\SI{}{\micro\second}$ distribution. The data points are:  MGPs (green circles), IGPs (orange squares), and all GPs (purple triangles).}
	\label{fig:pulseenergywidth}
\end{figure*}
\begin{figure*}
	\includegraphics[scale=0.405]{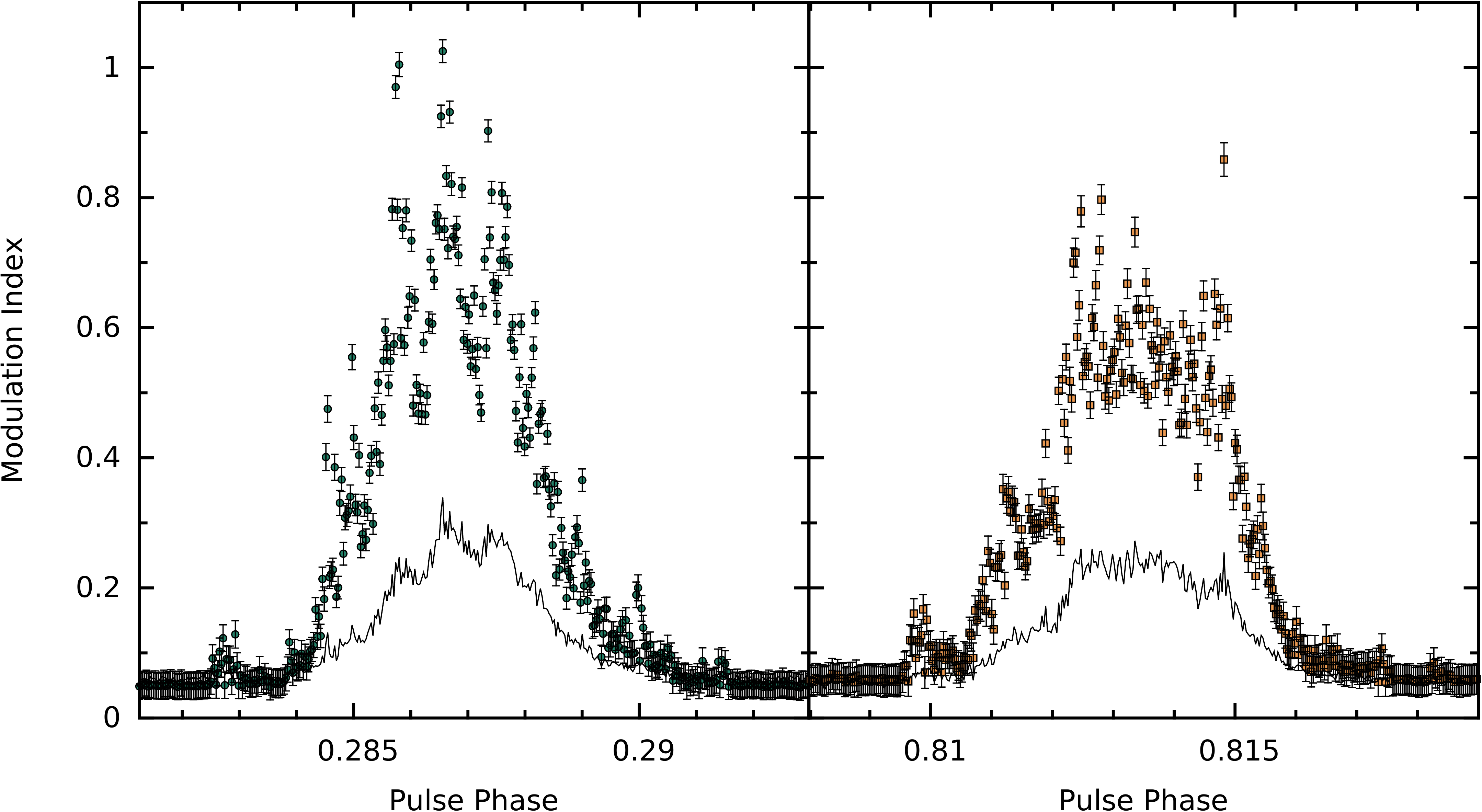}
	\centering
	\caption{Modulation indices ($m(\phi)$, points with error bars) for all MGPs (left) and IGPs (right), plotted together with the average intensity of the GPs (solid line). The contribution from the regular emission is visible at the leading edge of each profile. 
While the distributions appear to be approximately Gaussian, there is significantly more variation in the modulation indices around the centre of the profile ($\Delta m(\phi)\sim0.5$) when compared to the edges of the profile, in contrast to the regularity of non-GP single-pulse emission reported by \protect\citealp{jap01}.}
	\label{fig:GPmodulationindex}
\end{figure*}

\subsection{Flux Calibration} \label{fluxcalsection}
Our observations are not flux-calibrated, so we instead estimate the flux density $S$ of the GPs by using a modified version of the radiometer equation,
\begin{equation} \label{radiometerequation}
S=\frac{200k_{\text{B}}T_{\text{sys}}(\text{S/N)}}{A_{\text{e}}C\sqrt{W_{50}\Delta f}},
\end{equation}
where $k_{\text{B}}$ is the Boltzmann constant, $T_{\text{sys}}$ is the system temperature, $C$ is the coherency of the added observation, $A_{\text{e}}$ is the total collecting area of the telescopes used for the observation, $W_{50}$ is the full width at half-maximum of the GP, and $\Delta f$ is the bandwidth. We estimate a system temperature $T_{\text{sys}}=28$\,K, using typical values for the sky temperature at 1400\,MHz and receiver noise across the telescopes (3\,K and 25\,K respectively). 
The brightest GP we observed came from an observation on $22^{\text{nd}}$ May 2014 (MJD\,56799), and has an integrated signal-to-noise ratio of 1988 (Figure \ref{fig:brightGP}), corresponding to a pulse energy of 492\,Jy\,$\SI{}{\micro\second}$, or ${1.2\times10^{20}}$\,W at the distance of 3.27\,kpc estimated by \protect\cite{dcl+16}.
\begin{figure}
	\includegraphics[scale=0.3]{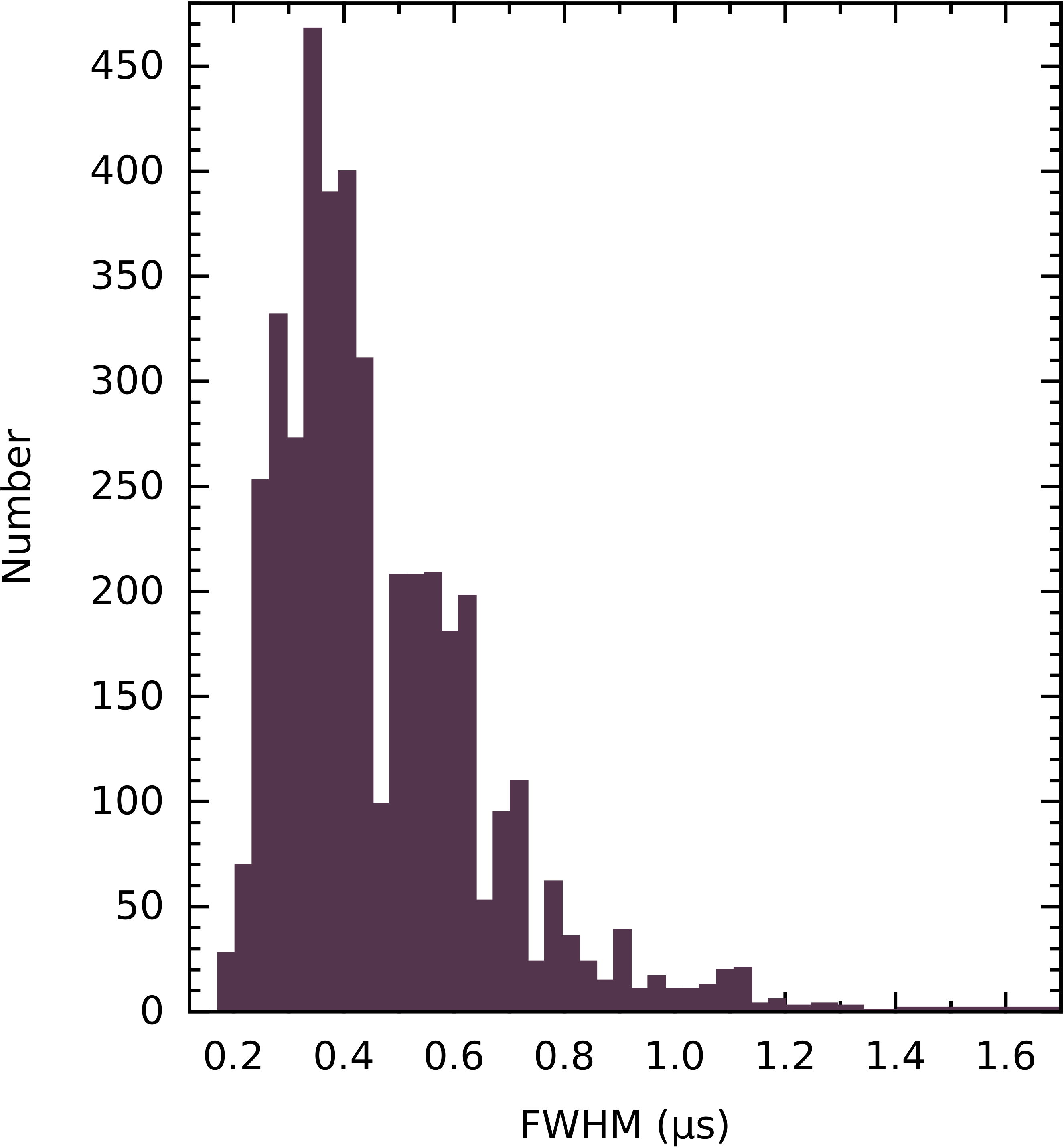}
\centering
	\caption{Distribution of observed GP pulse widths, measured at the half-maximum, after fitting an exponentially-modified Gaussian function to GPs that were smoothed with a Savitzky-Golay filter. The distribution has a clear peak at $\SI{0.35}{\micro\second}$, followed by a decay to more rare instances of broader pulse widths.}
	\label{fig:GPwidths}
\end{figure}
\begin{figure}
	\includegraphics[scale=0.355]{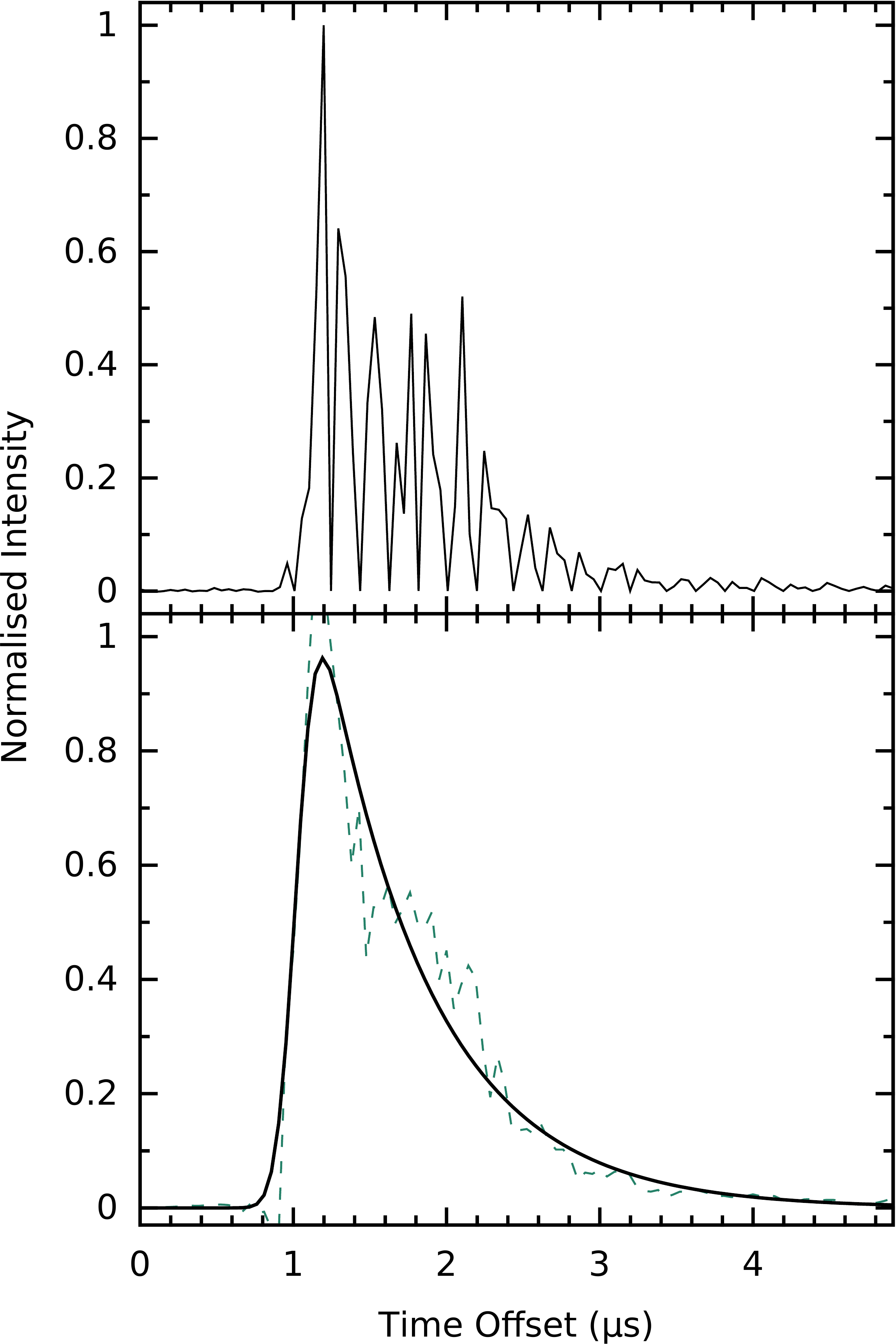}
	\centering
	\caption{Example of scattering observed in a bright GP. The profile (above) has a pulse energy of 242\,Jy\,$\SI{}{\micro\second}$ (integrated ${\text{S/N}=959}$) and a clear exponential scattering tail. The scattering time scale is measured (below) by fitting an exponentially-modified Gaussian function (solid line) to the observed GP (dashed line), after smoothing with a Savitzky-Golay filter (see text), which mitigates the effect of the `spiky' emission on the fitting process. The scattering time scale in this profile is measured to be $\tau_{\text{sc}}=676\pm32$\,ns. The separation of the visible quasi-periodic micro pulses is about 200\,ns, which is exemplary for the value determined for the larger sample of GPs of $249\pm197$\,ns (see Section \ref{gpemissionproperties} for details).}
	\label{fig:GPscatteringfit}
\end{figure}
\begin{figure*}
	\includegraphics[scale=0.335]{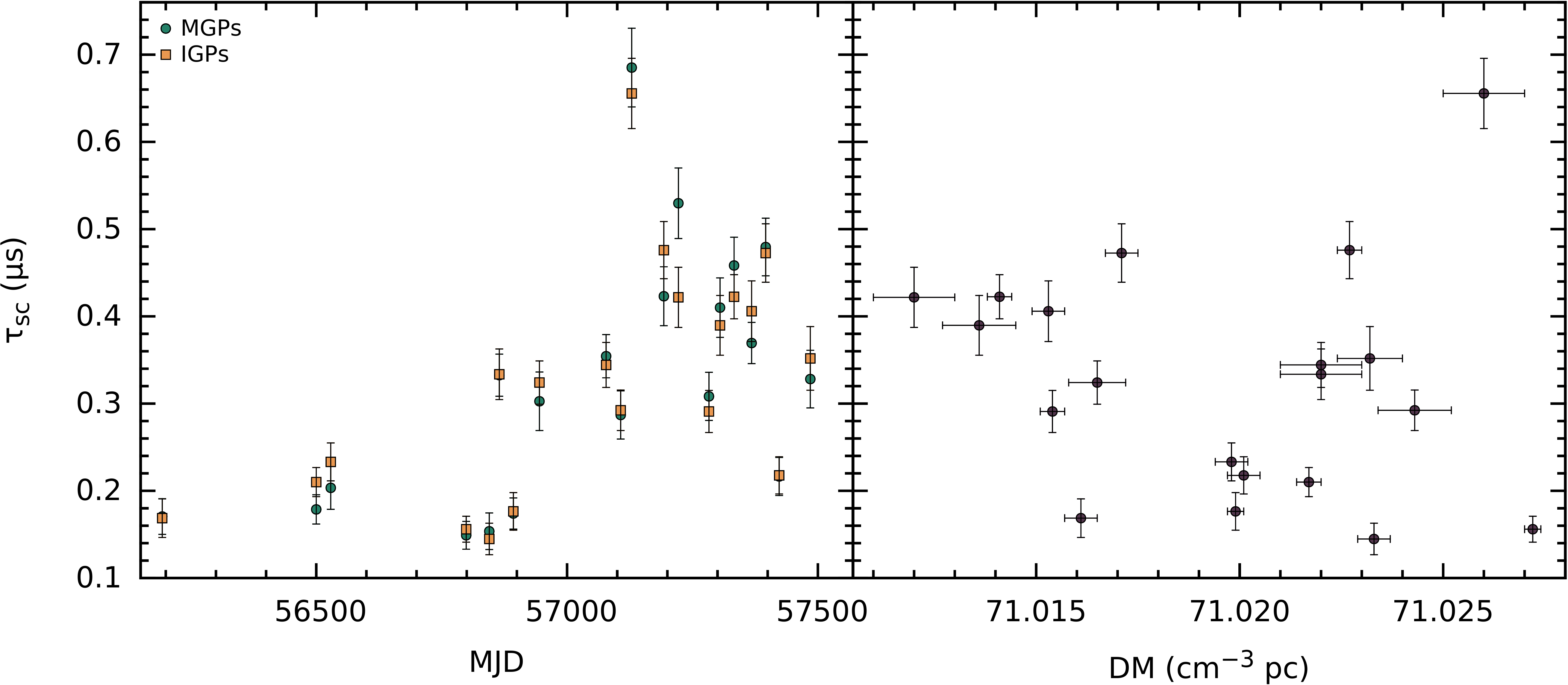}
	\centering
	\caption{\textit{Left:} Mean GP scattering time scales for each of our observations, measured using exponentially-modified Gaussian fits to each MGP (green circles) and IGP (orange squares) (see text for details). The measurements from each GP component are in good agreement and show significant variation across our observations. \textit{Right:} Mean scattering vs. DM for each of our observations. The two quantities do not appear to be correlated, in contrast to the strong correlation observed in pulsars such as the Crab.}
	\label{fig:GPmeanscattering}
\end{figure*}

\subsection{Giant Pulse Energy Distributions} \label{pulseenergysection}
We plot the cumulative pulse energy distributions of the GPs in our sample in Figure \ref{fig:gp_flux_numbers}, and observe the expected power law distribution for high pulse energies. 
The pulse energy distribution turns over at low energies of $\lesssim2$\,Jy\,$\SI{}{\micro\second}$, which is expected due to the finite number of GPs emitted, and as not every rotation contains a GP. 
While the pulse energies of the IGP distribution are well-modelled with a single power law, we find that the MGP and all-GP distributions are better-modelled with a broken power law, with pulse energies $\gtrsim7$\,Jy\,$\SI{}{\micro\second}$ following a flatter distribution in a similar way to high-energy GPs from the Crab Pulsar \protect\citep{ps07}. 
We do not observe any pulse-width-dependent variation in the pulse energy at which the power law break occurs (Figure \ref{fig:pulseenergywidth}), in contrast to that observed in the pulse energy distribution of the Crab Pulsar by \protect\cite{ps07}.
The power law indices we measure are summarised in Table \ref{tab:powerlawindices}, and differ from those reported by \protect\cite{spb+04} and \protect\cite{cst+96}, which are $\alpha=-1.40\pm0.01$ and $\alpha=-1.8\pm0.1$ respectively for the all-GP distribution. We note that these values were determined using sample sizes approximately one order of magnitude lower than our sample size, and that somewhat different observing frequencies were used. We note also that the power law indices for the distributions of our individual observations vary significantly, ranging from $\alpha=-1.4\pm0.1$ to $\alpha=-5.8\pm0.4$ for the all-GP distributions. Drawing random MGPs and IGPs from our distribution, to form sample sizes equal to that of \protect\cite{spb+04} (309 GPs), we find a mean all-GP power law index of $\alpha=-3.9\pm0.1$ after 1000 trials.

\begin{figure}
	\includegraphics[scale=0.35]{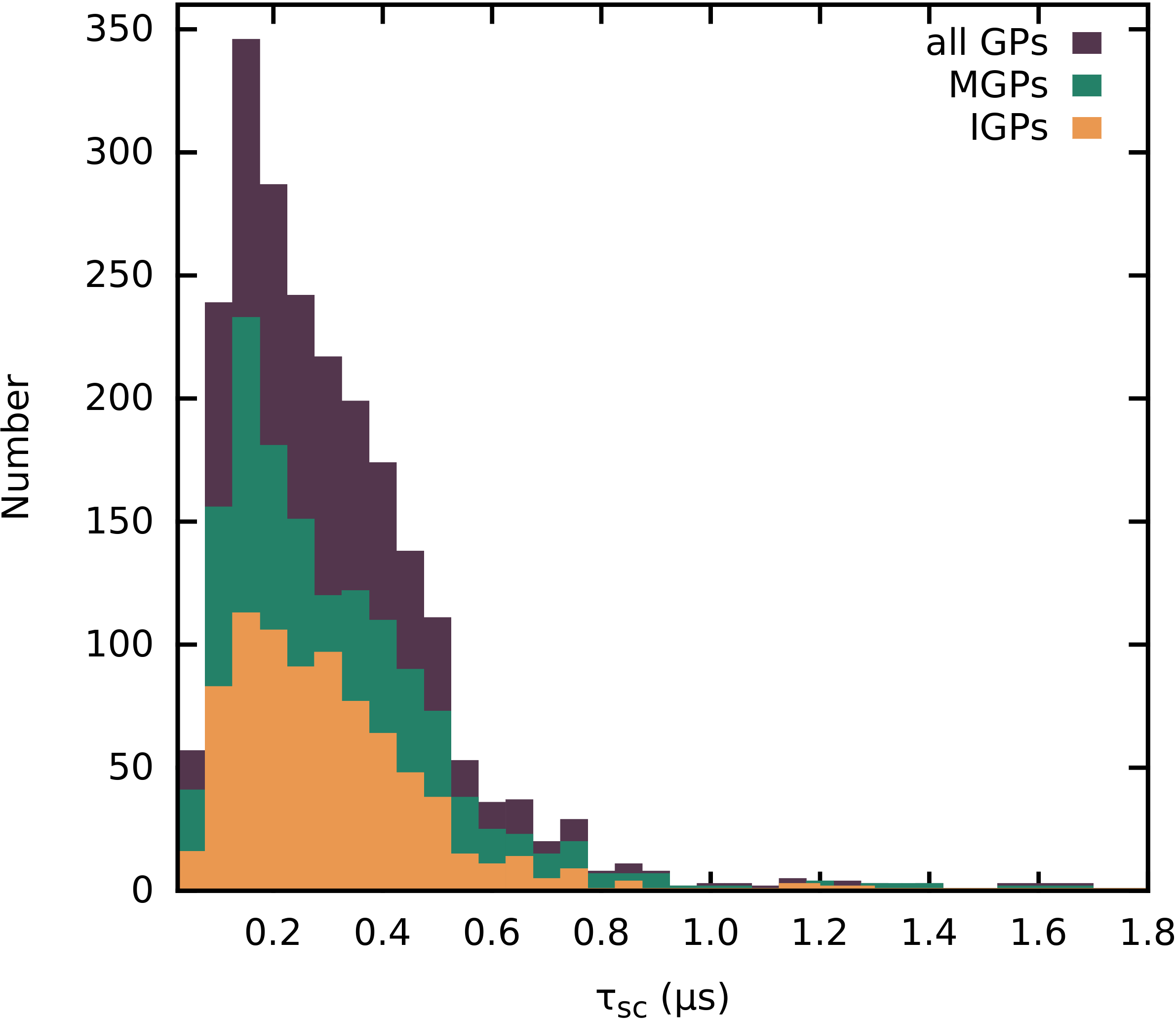}
	\centering
	\caption{Scattering time scales measured from the exponentially-modified Gaussian fit to the smoothed GPs. The measured values are binned at intervals of 50\,ns, and the most common time scale is $\SI{0.15}{\micro\second}$. The time resolution of our observations does not allow scattering time scales $<60$\,ns to be identified.}
	\label{fig:GPscatteringhistogram}
\end{figure}

\subsection{Modulation Indices} \label{modulationindexsection}
The intensity fluctuations between individual pulses can be quantified by the phase-resolved modulation index given by (e.g. \protect\citealp{jg04}, \protect\citealp{wes06})
\begin{equation} \label{modulationindexequation}
m(\phi)=\frac{\sqrt{\langle I(\phi)^{2}\rangle-\langle I(\phi)\rangle^{2}}}{\langle I(\phi)\rangle},
\end{equation}
where $I$ is the intensity, $\phi$ is the pulse phase, and the angle brackets indicate averaging. A GP pulse stack (i.e. total intensity as a function of pulse phase for successive GPs) was generated for each observation, and the modulation indices were computed for the MGP and IGP separately, using the \textsc{PSRSALSA}\footnote{\url{github.com/weltevrede/psrsalsa}} suite \protect\citep{wel16}. The resulting plots were aligned and averaged using the phase information of the average pulse profile (Figure \ref{fig:GPmodulationindex}). We find that the modulation index distribution as a function of pulse phase is approximately Gaussian, and that the modulation indices vary by $\sim50\%$ around the peak of both the IGP and MGP pulse phase distributions, while the edges of the distributions do not vary significantly. This is different to the variations seen in most pulsars for which modulation indices are measured, which are usually observed to be greater at the profile edges than at the centre (e.g. \protect\citealp{cal+13}). We note that in Figure \ref{fig:GPmodulationindex}, a significant increase in modulation index, which appears to precede both GP regions, is visible. This increase may indicate that modulation is present in the regular pulse region, in contrast to the findings of \protect\cite{jap01}, where intensity modulation in observations taken at 430\,MHz was found to be consistent with diffractive interstellar scattering and scintillation . Further investigation is required to determine whether the effect is truly related to the regular emission. 

\begin{figure}
	\includegraphics[scale=0.36]{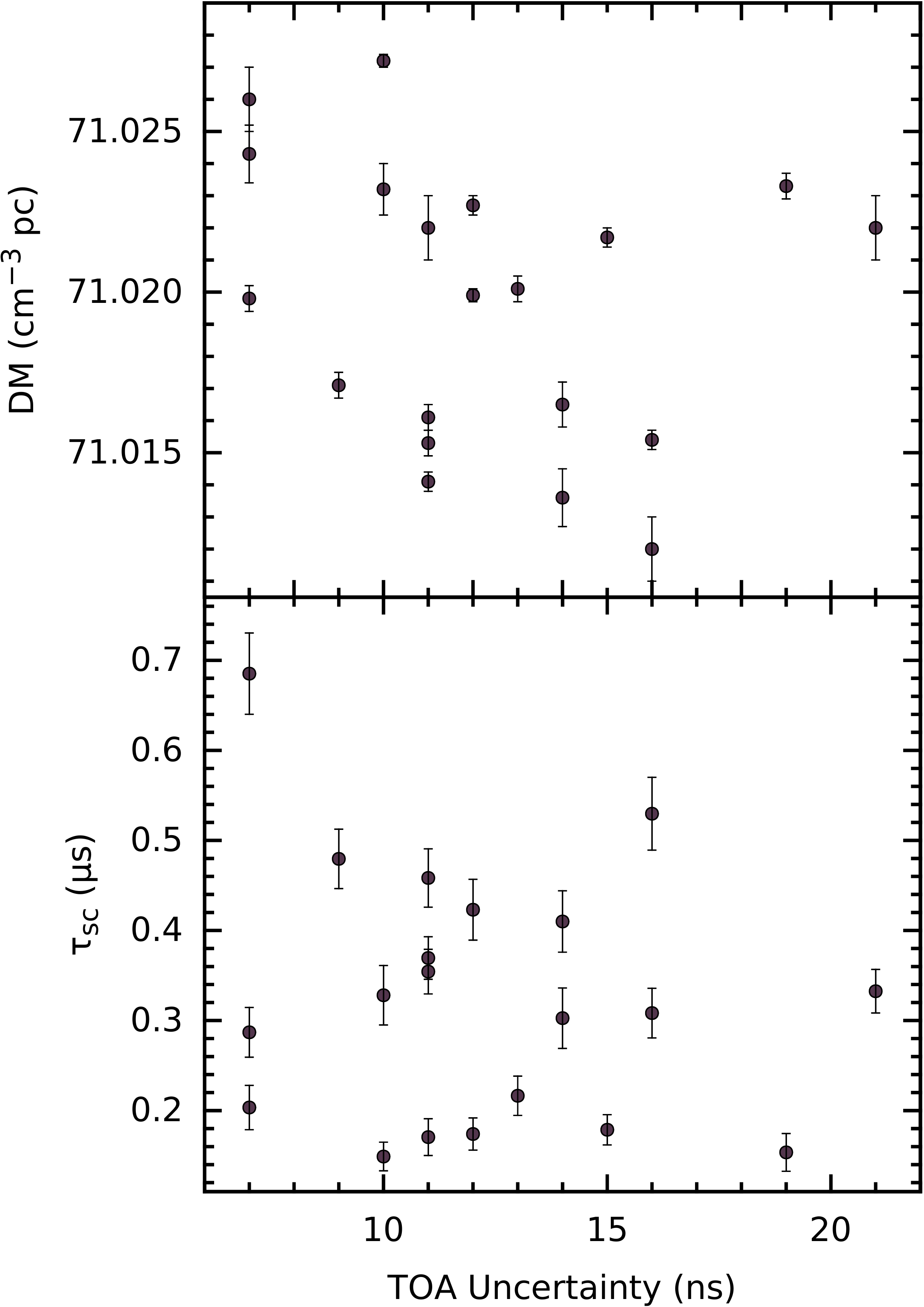}
	\centering
	\caption{Mean scattering time scales measured from the GPs (below), and the mean DM measured from GPs (above) vs. TOA uncertainty measured using the regular pulse profile. In both plots, the two quantities do not appear to be correlated.}
	\label{fig:GPtoaerrorscatteringdm}
\end{figure}
\begin{figure}
	\includegraphics[scale=0.35]{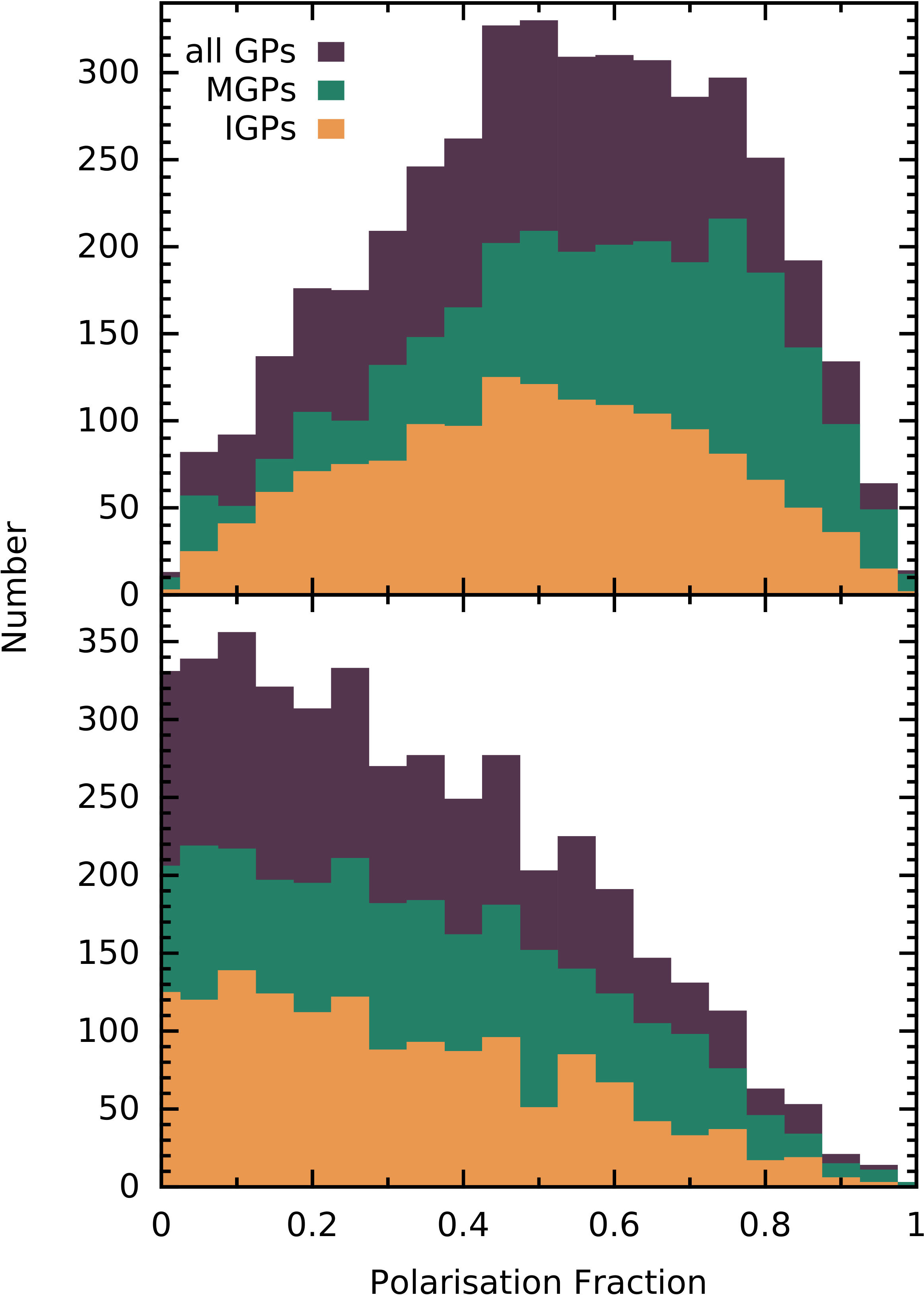}
	\centering
	\caption{Histograms of the fractions of linear (L/I, above) and circular (|V|/I, below) polarisation of the MGPs (green), IGPs (orange), and all GPs (purple) in our data set.}
	\label{fig:GPpolarisationhistogram}
\end{figure}
\subsection{Scattering and DM Variations} \label{scatteringsection}
The measured DM of PSR\,B1937+21 is known to fluctuate significantly between observations (\protect\citealp{cwd+90}, \protect\citealp{rdb+06}), and it has been suggested that scatter-broadening is the only cause of variations in the observed pulse shape of the regular emission \protect\citep{jap01}.
We observe a wide distribution of GP widths (Figure \ref{fig:GPwidths}), similar to the findings of \protect\cite{ksv10} of the GP width variations due to large changes in scattering in the Crab Pulsar.
Our data set allowed us to make many separate scattering measurements over the course of each observation, using individual GPs. As observed GPs are often composed of several individual peaks, we smooth the pulse shape using a Savitzky-Golay filter \protect\citep{sg64}, enabling the smoothed pulse to be approximated as an exponentially-modified Gaussian (Figure \ref{fig:GPscatteringfit}, and see \protect\citealp{mck14}, \protect\citealp{mls+18}). The scattering time scales measured using an exponentially-modified Gaussian fit to the smoothed IGP and MGP for each observation are in good agreement (Figure \ref{fig:GPmeanscattering}).

The distribution of our measured scattering time scales presented in Figure \ref{fig:GPscatteringhistogram} is in good agreement with those obtained by \protect\cite{cwd+90} at observing frequencies of 430\,MHz, assuming that scattering scales with observing frequency as $\tau_{\text{sc}}\propto f^{-4.4}$ (the frequency scaling of scattering time scales in PSR\,B1937+21 between 400\,MHz and 2\,GHz has been found by \protect\citealp{kps+07b} to be consistent with the expected $f^{-4.4}$ scaling). The distribution has a clear peak at a time scale $\tau_{\text{sc}}=\SI{0.15}{\micro\second}$, and $\sim64\%$ of pulses have scattering time scales $<\SI{0.35}{\micro\second}$ . As expected from the findings of \protect\cite{rdb+06}, the mean scattering time scale was found to vary significantly between observations (Figure \ref{fig:GPmeanscattering}), and instances of very high scattering were found to skew the mean scattering time scales for the observations.

We find no correlation between the mean scattering time scale and DM (Figure \ref{fig:GPmeanscattering}) measured in each of our observations. 
This is in contrast to the strong correlation observed between scattering time scales and DM seen in the Crab Pulsar (\protect\citealp{klj+08}, \protect\citealp{mls+18}).
We also find no correlation between the mean scattering time scale and DM, and the normal pulse time of arrival (TOA) error (Figure \ref{fig:GPtoaerrorscatteringdm}). This indicates that unmodelled variations in these quantities do not limit the precision of our TOAs for PSR\,B1937+21, at our observing frequencies and sensitivity.

\begin{figure*}
	\includegraphics[scale=0.27]{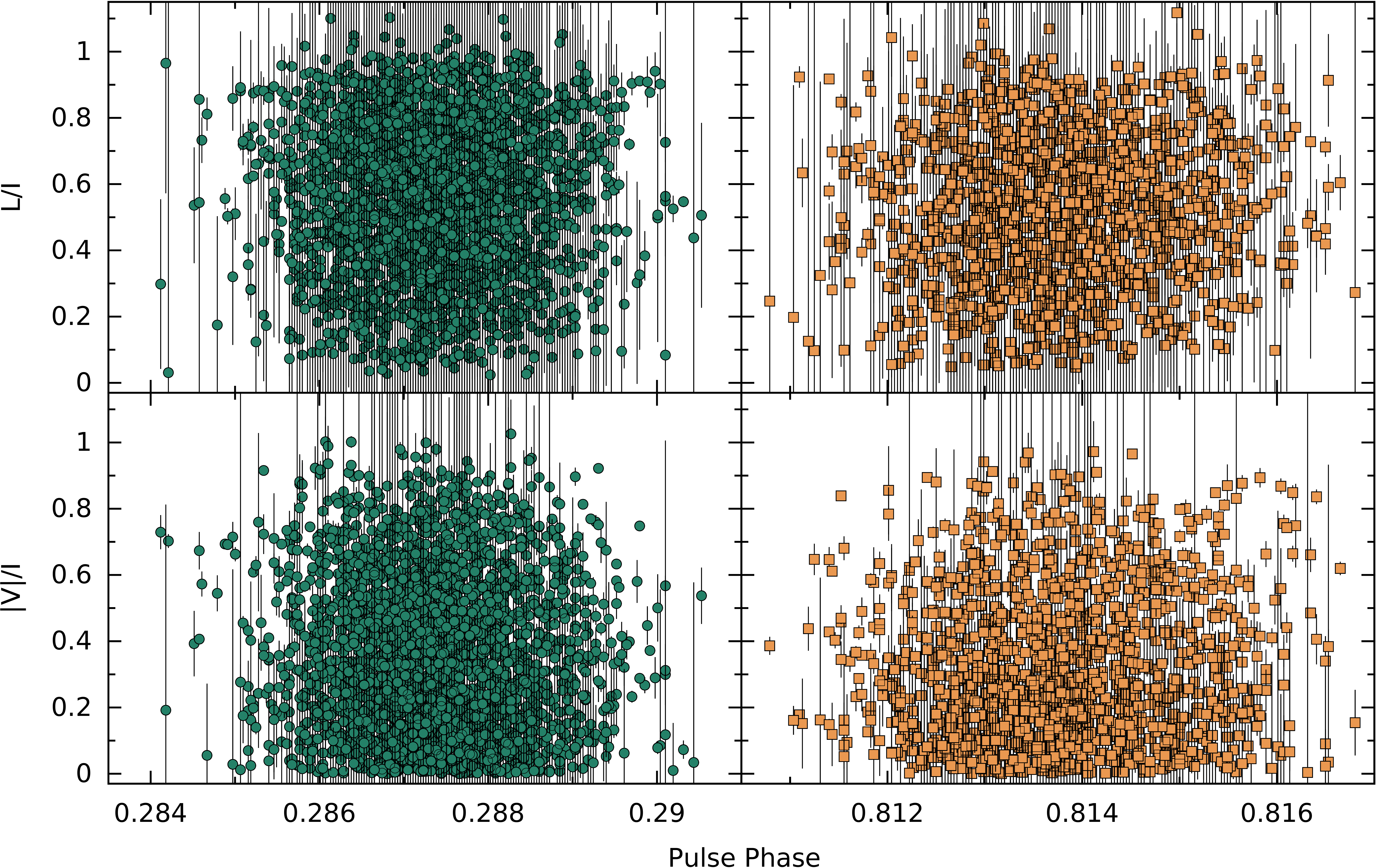}
	\caption{Distributions of fractional linear (L/I, top row) and circular (|V|/I, bottom row) polarisation of MGPs (left column, green) and IGPs (right column, orange), according to their phase location relative to the peak MP total-intensity emission (centred at a phase of 0.25). The fractional polarisation of the emission does not appear to be dependent on the location within the emission phase, nor are there significant differences between the MGP and IGP distributions.}
	\label{fig:GPpolarisationphase}
\end{figure*}

\subsection{Giant Pulse Polarisation} \label{polarisationsection}
While the emission from the average pulse profile of PSR\,B1937+21 is approximately $50\%$ linearly polarised and has very little circular polarisation (Figure \ref{fig:B1937polprof}), GPs from PSR\,B1937+21 are often observed to be highly polarised, with some bright GPs displaying almost $100\%$ circular polarisation (\protect\citealp{cst+96}, \protect\citealp{spb+04}). The GPs in our sample are not rotation measure corrected. However, we estimate that the error introduced to the position angle (PA) between the top and bottom of our maximum bandwidth is $\sim0.07$\,rad, for a rotation measure of 8.3\,rad\,m$^{-2}$ \protect\citep{dhm+15}, which will not significantly alter the polarisation properties. 
The polarisation profile of the GPs averaged over all 21 observations is different to that of the averaged regular pulse profile (Figure \ref{fig:averageprofile}), implying that the polarised-emission statistics of GPs are different to those of the regular individual pulses. 
There are some difficulties in measuring polarised emission at time resolutions approximately equal to the Nyquist sampling rate, as is the case for our data set, or in cases where the sampling resolution is approximately equal to the scattering time scale. In these cases, unresolved emission can appear to be $100\%$ polarised in each sample (\protect\citealp{cor76}, \protect\citealp{van09}), although in our analysis we average over multiple samples and frequency channels, and thus our polarisation measurements are not impacted by our choice of sampling resolution. 
In Figure \ref{fig:GPpolarisationhistogram}, we plot histograms of the observed fractions of linear and circular polarisation from our GPs. 
We find that the GPs in our data set tend to be highly linearly-polarised, and we observe that while there are instances of almost entirely-polarised GPs, these are comparatively rare. Of the GPs in our data set, $1.1\%$ of MGPs and $0.6\%$ of IGPs display $>90\%$ circular polarisation, while $6.0\%$ of MGPs and $3.9\%$ of IGPs are $>90\%$ linearly polarised. We find that while the histograms of fractional circular polarisation are similar for MGPs and IGPs, the fractional linear polarisation histograms are different for the highly-polarised MGPs and IGPs, with a higher portion of MGPs displaying $>50\%$ linear polarisation. 
We find no evidence for phase-dependence of polarised emission, nor are there significant differences between the IGP and MGP polarisation fraction distributions (Figure \ref{fig:GPpolarisationphase}), in agreement with the findings of \protect\cite{zps+13}.
We also find no correlation between GP pulse energy and fractional polarisation or GP width.

\begin{table*}
\large
\caption{Pulsars for which bright single-pulse emission has been detected, and the corresponding magnetic field strength at the light cylinder ($B_{\text{LC}}$). The `rank' column refers to the position of the pulsar when sorting the 2109 pulsars for which this property is currently listed in the ATNF Pulsar Catalogue by their magnetic field strengths at the light cylinder. Pulsars which are classified as GP-emitters have $B_{\text{LC}}>10^{5}$\,G (see text).}
\label{tab:GPemitters}
\centering 
\begin{tabular} {l c c c}
\hline 
Name & $B_{\text{LC}}$ (G) & Rank & Reference\\
\hline
\hline
B1937+21 (J1939+2134) & $1.0\times10^{6}$ & 2 & \protect\cite{wcs84}, \protect\cite{cst+96} \\
B0531+21 (J0534+2200) & $9.6\times10^{5}$ & 3 & \protect\cite{sr68} \\
B1821$-$24A (J1824$-$2452A) & $7.4\times10^{5}$ & 4 & \protect\cite{rj01} \\
B1957+20 (J1959+2048) & $3.8\times10^{5}$ & 6 & \protect\cite{jkl+04} \\
B0540$-$69 (J0540$-$6916) & $3.6\times10^{5}$ & 7 & \protect\cite{jr03} \\
J0218+4232 & $3.2\times10^{5}$ & 9 & \protect\cite{jkl+04} \\
B1820$-$30A (J1823$-$3021A) & $2.5\times10^{5}$ & 16 & \protect\cite{kbm+05}  \\
\hline
B0656+14 (J0659+1414) & 770 & 437 & \protect\cite{ke06}  \\
B0950+08 (J0953+0755) & 140 & 790 & \protect\cite{sin01}, \protect\cite{smi12} \\
J1752+2359 & 71 & 1007 & \protect\cite{ek05}  \\
B0529$-$66 (J0529$-$6652) & 39 & 1183 & \protect\cite{cal+13} \\
B0031$-$07 (J0034$-$0721) & 7.0 & 1694 & \protect\cite{kel04} \\
B1112+50 (J1115+5030) & 4.2 & 1832 & \protect\cite{ek03} \\
B1237+25 (J1239+2453) & 4.1 & 1843 & \protect\cite{kp17} \\
\hline
\end{tabular}
\end{table*}

\section{Discussion} \label{discussion_section}
\subsection{Giant Pulse Emission Properties} \label{gpemissionproperties}
The phase window in which GPs occur may allow the properties of the GP emission region to be constrained. For example, a model by \protect\cite{lyu07} describes GPs in the Crab Pulsar as originating from the last closed field lines, in contrast to the regular emission, which originates from the open field lines. In the \protect\cite{lyu07} model, the observer angle to this section of the field lines must fall within a narrow ($\sim10^{-3}$\,rad) range, implying that the rarity of known GP-emitting pulsars is due to the low probability that this specific alignment will occur, and could explain the narrow phase region in which GPs are observed (Figure \ref{fig:averageprofile}). 

We use the definition of GPs as i) possessing pulse energies $>10$ times the single-pulse average ii) being much narrower in pulse width than the average pulse profile iii) being confined to a narrow phase window. The pulsars for which GP emission has been detected exhibit some of the highest-known magnetic field strengths at the light cylinder ($>10^{5}$\,G, summarised in Table \ref{tab:GPemitters}), given by (e.g. \citealp{lk05})
\begin{equation} \label{bfieldatlc}
{B_{\text{LC}}\sim3\times10^{8}\left(\dot{P}^{1/2}P^{-5/2}\right)} \ , 
\end{equation}
and in fact six of the ten pulsars with the highest-measured $B_{\text{LC}}$ are GP emitters. It has therefore been proposed that a high $B_{\text{LC}}$ is potentially a crucial discriminator for GP emission \protect\citep{cst+96}. Of the other pulsars which place in the top 10 highest $B_{\text{LC}}$, only the radio-quiet X-ray pulsar J0537$-$6910 (which possesses the highest-known $B_{\text{LC}}$) has been searched for GPs, with none detected in a 12 hour observation at 1390\,MHz \citep{cmj+05}. 
It should be noted that bright single pulses have been detected from pulsars with much lower $B_{\text{LC}}$ ($<1000$\,G), but due to the broader pulse widths, and in some cases the wider phase window in which the bright pulses occur, we do not classify these pulsars as GP-emitters.

\begin{figure*}
	\includegraphics[scale=0.998]{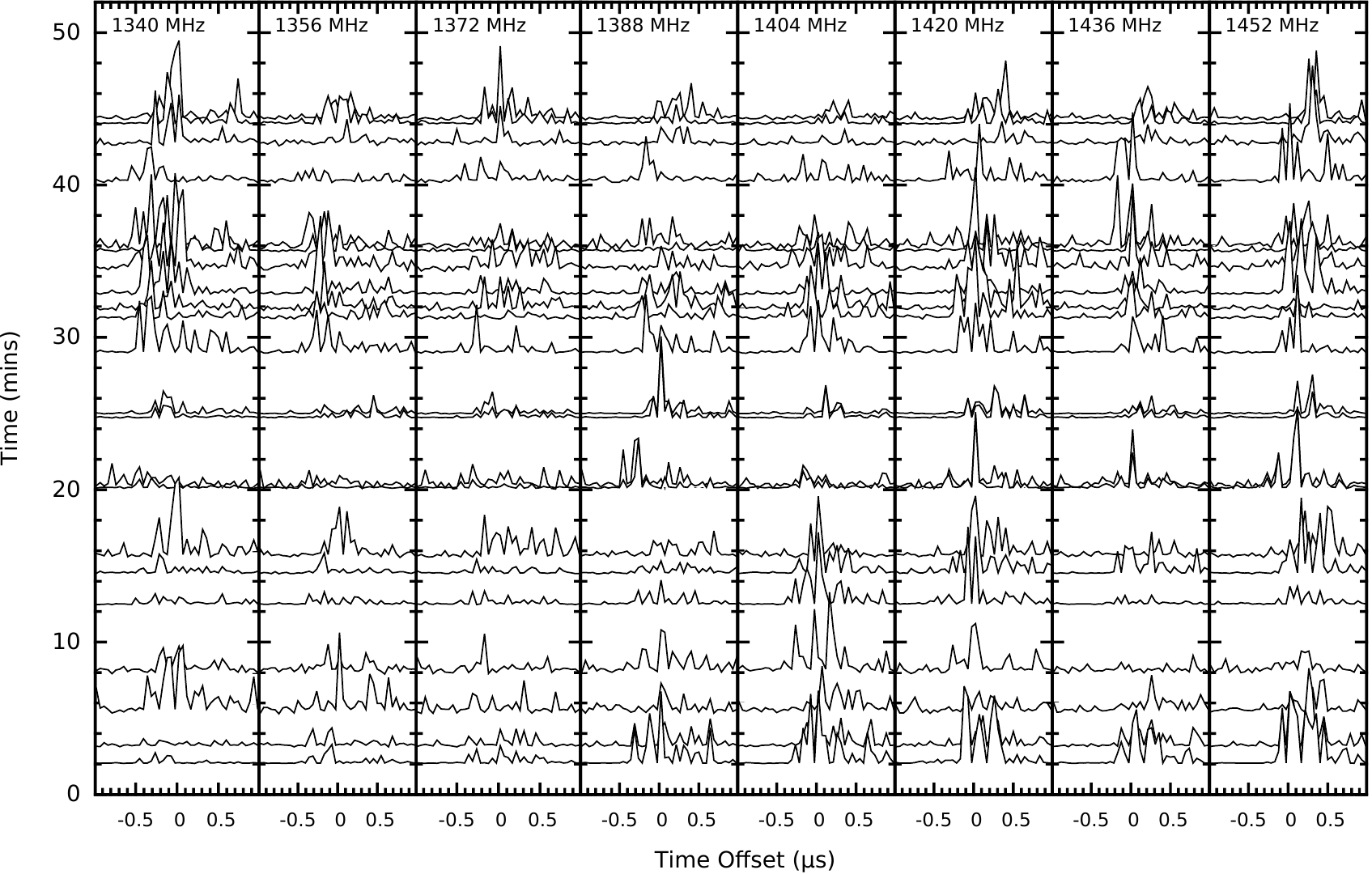}
\end{figure*}
\begin{figure*}
\hspace*{-0.03cm}  
	\includegraphics[scale=1.0]{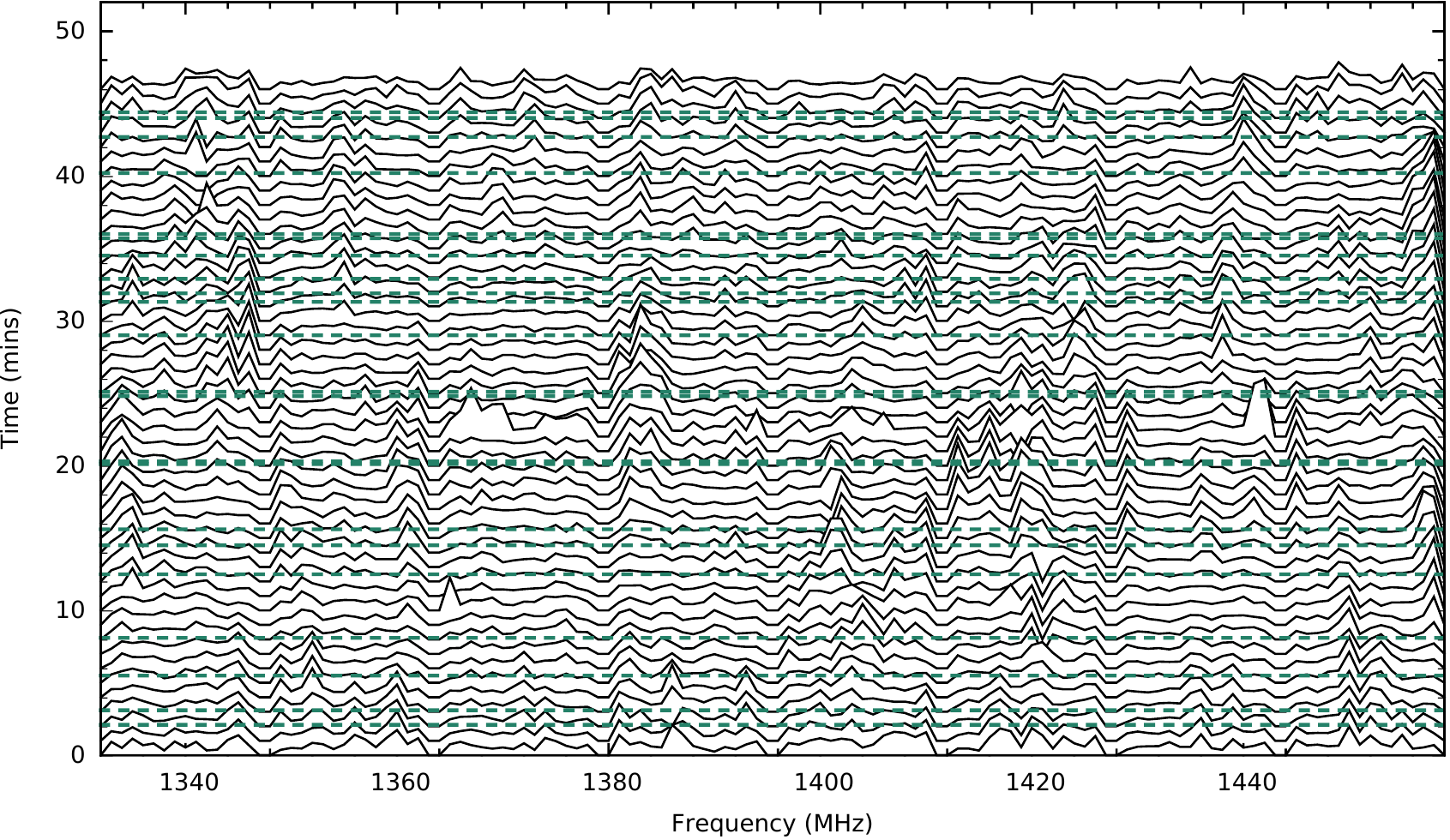}
	\centering
	\caption{Comparison of frequency-resolved GP shapes with the full-observation dynamic spectrum, using data from 7$^{\text{th}}$ November 2015 (MJD\,57333). \textit{Above}: Total-intensity MGPs, plotted by sub-band. Each GP is normalised by the highest intensity across all frequencies, and centred on the phase of the highest-intensity bin in the frequency-averaged GP profile. Only GPs with $\text{S/N}>8\sigma$ are plotted, to preserve clarity. \textit{Below}: Dynamic spectrum of the total observation, averaged over 1-min sub-integrations, and normalised by the highest intensity across the total observation. The nulls at the band edges are due to masking of instrumental artefacts. The dashed lines correspond to the GPs in the above panel.}
	\label{fig:gp_scint_structure_MP}
\end{figure*}
\begin{figure*}
	\includegraphics[scale=0.998]{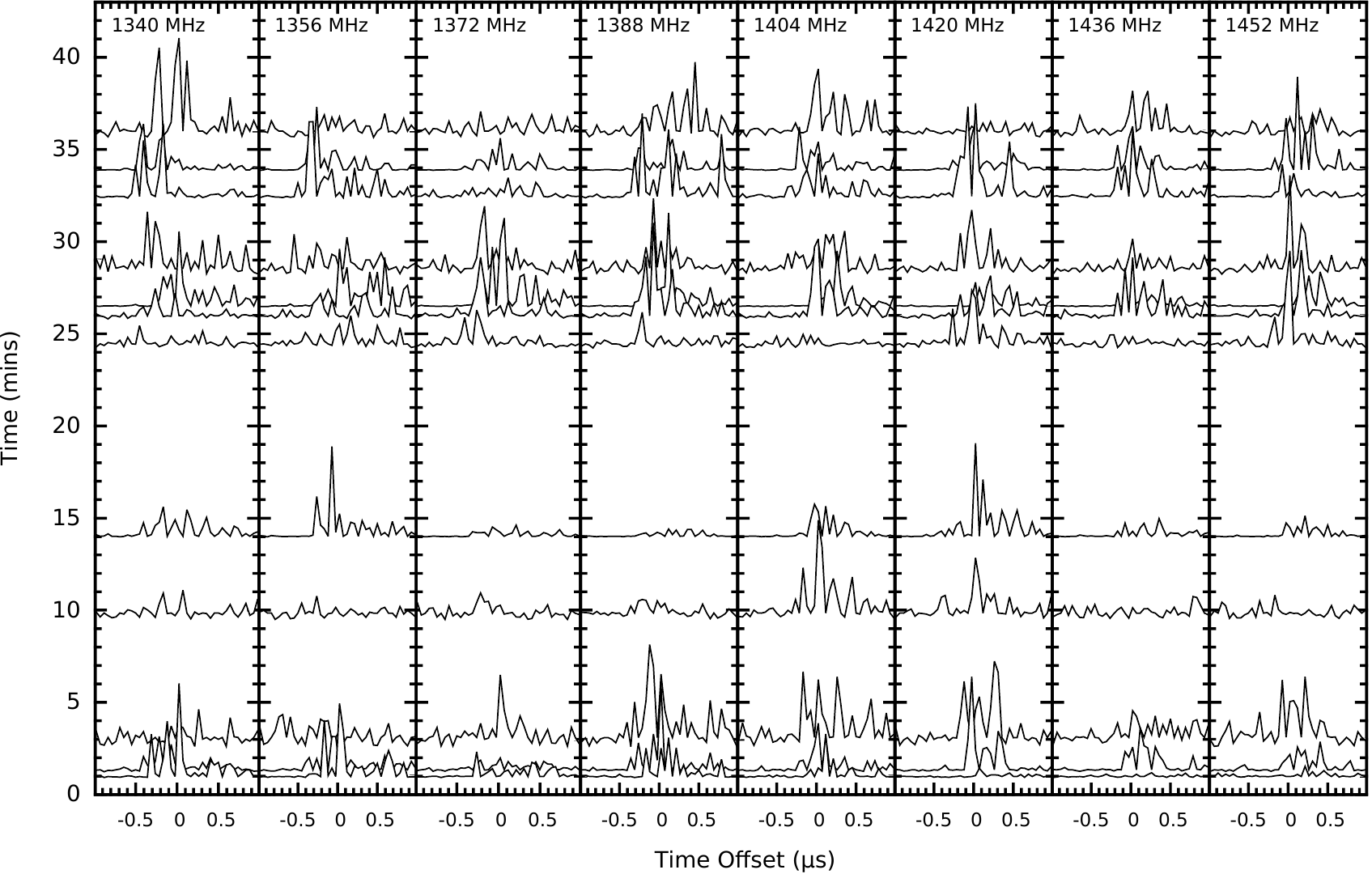}
\end{figure*}
\begin{figure*}
\hspace*{-0.03cm}    
	\includegraphics[scale=1.0]{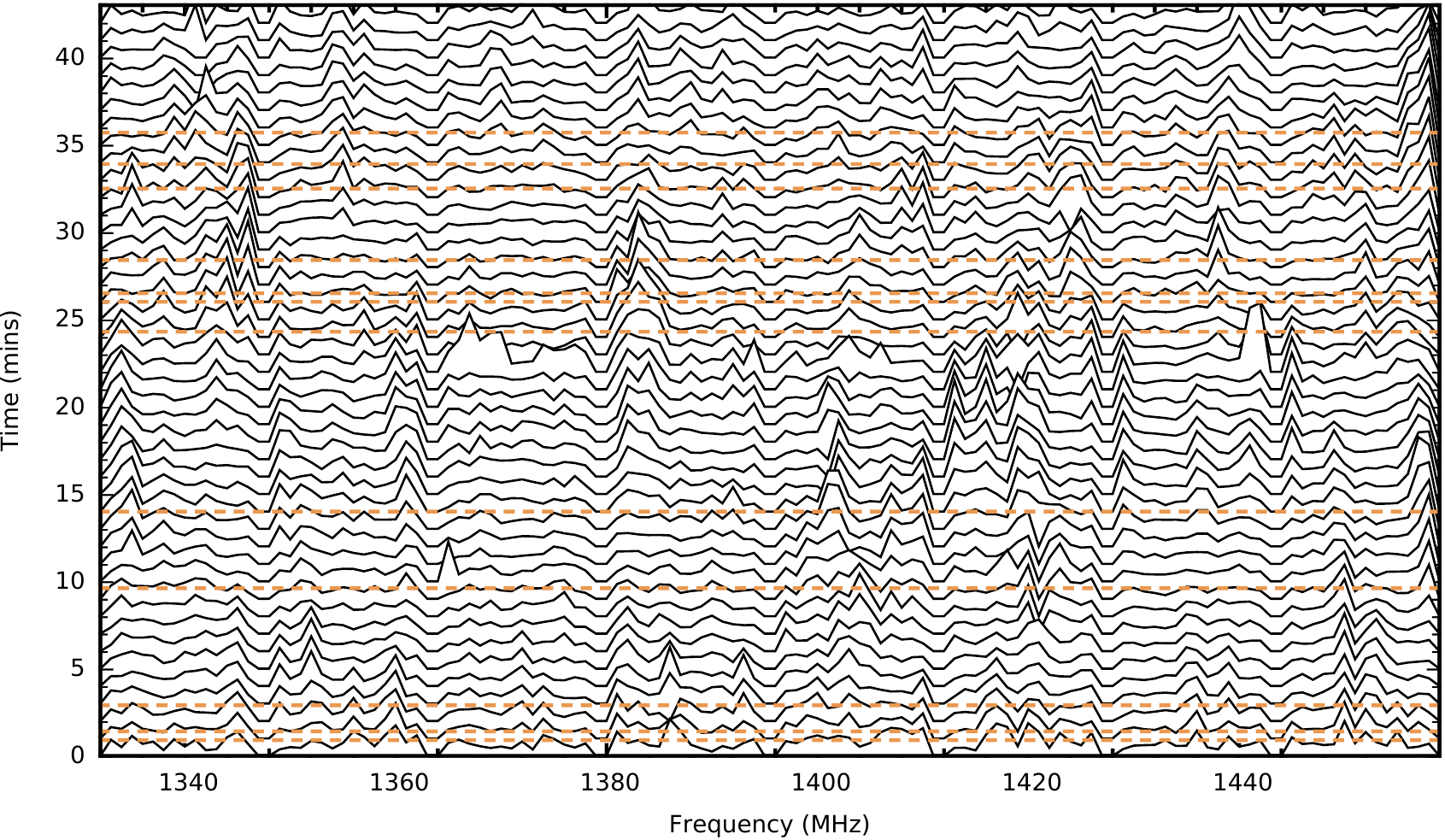}
	\centering
	\caption{As Figure \ref{fig:gp_scint_structure_MP}, for the IGPs.}
	\label{fig:gp_scint_structure_IP}
\end{figure*}

\begin{figure*}
	\includegraphics[scale=1.0]{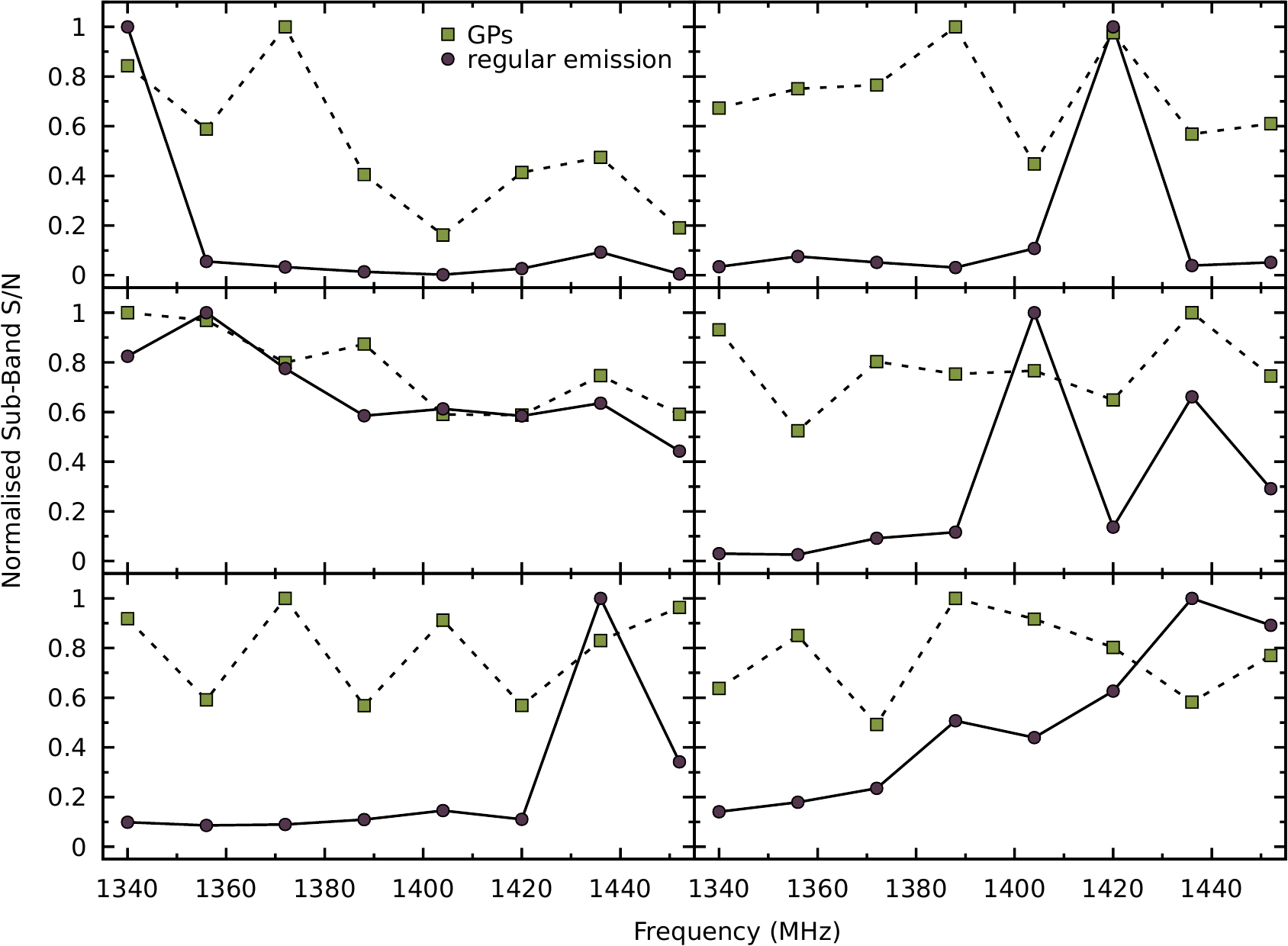}
	\centering
	\caption{Comparison of sub-band S/N for selected bright GPs (purple circles, dashed lines) and the regular emission integrated over 10 seconds and coincident with the GP (yellow squares, solid lines). While there may be some correlation between the GP and non-GP spectra, there is underlying intrinsic frequency dependence on the brightness of the GP emission.}
	\label{fig:gp_scint_structure_compare}
\end{figure*}

As mentioned earlier, GPs are believed to be a radio component of the high-energy emission, and they may be expected to originate from a different altitude or location in the magnetosphere to that of the regular radio emission. A widely-used method for measuring the emission height of different components of the pulsar beam is given by \protect\cite{bcw91}. In their model, the phase at which the steepest gradient of the polarisation PA occurs is interpreted as the closest approach of the line of sight to the pulsar's magnetic pole, and the emission height is calculated using the phase difference between this fiducial point and the peak intensity of the pulse profile. For this method to be used, the precise polarisation PA morphology is required. However, GPs from PSR\,B1937+21 are observed to occur at the edge of the regular emission region i.e. widely separated from the magnetic pole, and where the polarisation PA is found to flatten. Additionally, scattering by the interstellar medium, which is known to be relatively high in PSR\,B1937+21 (e.g. \protect\citealp{lmj+16}), also flattens the polarisation PA \citep{lh03}. These two effects give rise to a featureless GP polarisation PA, making the \protect\cite{bcw91} model difficult to apply to measuring the GP emission height. The emission height of GPs has also been measured (e.g. \protect\citealp{kss11}) using a model by \protect\cite{knn+82}, where the frequency-dependence of the component separation is used to determine the difference in emission heights. However, as MSP emission heights exhibit very little variation with frequency \protect\citep{kll+99}, a study like this would require a much wider bandwidth than our maximum of 128\,MHz, or simultaneous multi-frequency observations of GPs.

The GPs in our data set vary significantly in shape and intensity over the observing bandwidth throughout each observation. In Figure \ref{fig:gp_scint_structure_MP} and Figure \ref{fig:gp_scint_structure_IP}, we plot the 22 highest-S/N GPs from the observation from 7$^{\text{th}}$ November 2015 (MJD\,57333, chosen due to its high coherency use of the full 128\,MHz bandwidth), separated into eight 16-MHz sub-bands. We observe that both the MGPs and IGPs are composed of multiple components with widths smaller than our time resolution, and that these components vary in relative intensity with frequency and time of emission. This variation in component intensity with frequency is not coincident with the scintillation at the time of emission, and must therefore be an intrinsic property of the emission. When comparing the sub-band S/N for GPs with the regular emission (Figure \ref{fig:gp_scint_structure_compare}), there is a slight correlation, which may be expected if in this case scintillation is affecting the brightness of the GPs, as noted by \cite{ps03}. However, very bright emission is observed in single bands of most GPs, which cannot be accounted for by scintillation alone, and indicates that there is intrinsic variability in the GP emission spectra. Additionally, the typical scintillation bandwidth, estimated from the typical scattering time scales presented in Figure \ref{fig:GPscatteringhistogram}, is $\sim7$\,MHz, or less than half of the width of the sub-bands we have used. 

\begin{figure}
	\includegraphics[scale=0.36]{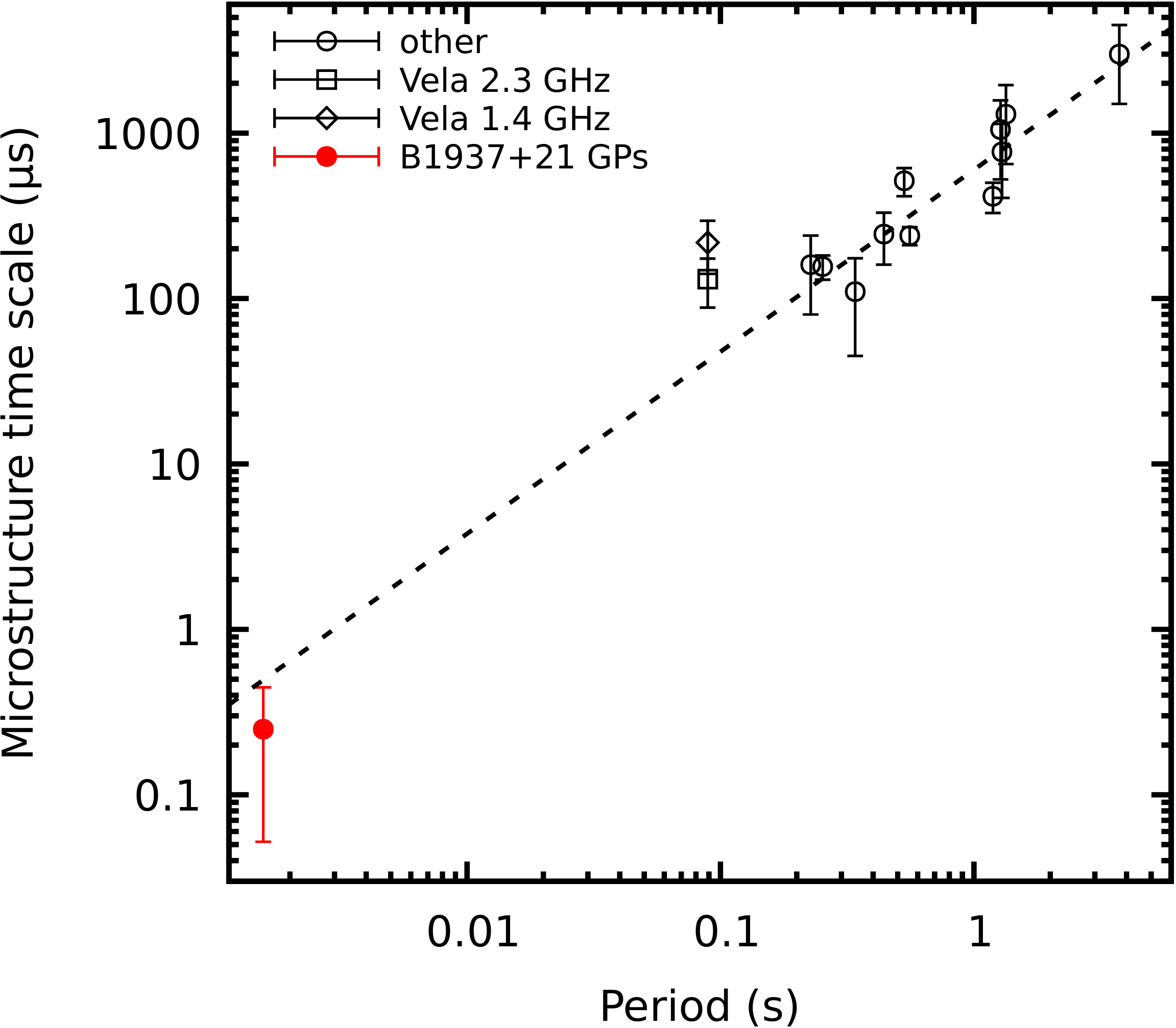}
	\centering
	\caption{PSR\,B1937+21 microstructure time scale used to update Figure 4 of \protect\cite{kjv02}. The dashed line represents the best-fit derived by \protect\cite{kjv02}, which is in good agreement with the time scale we measure for the GPs in our data set.}
	\label{fig:microscale}
\end{figure}
\begin{figure*}
	\includegraphics[scale=0.26]{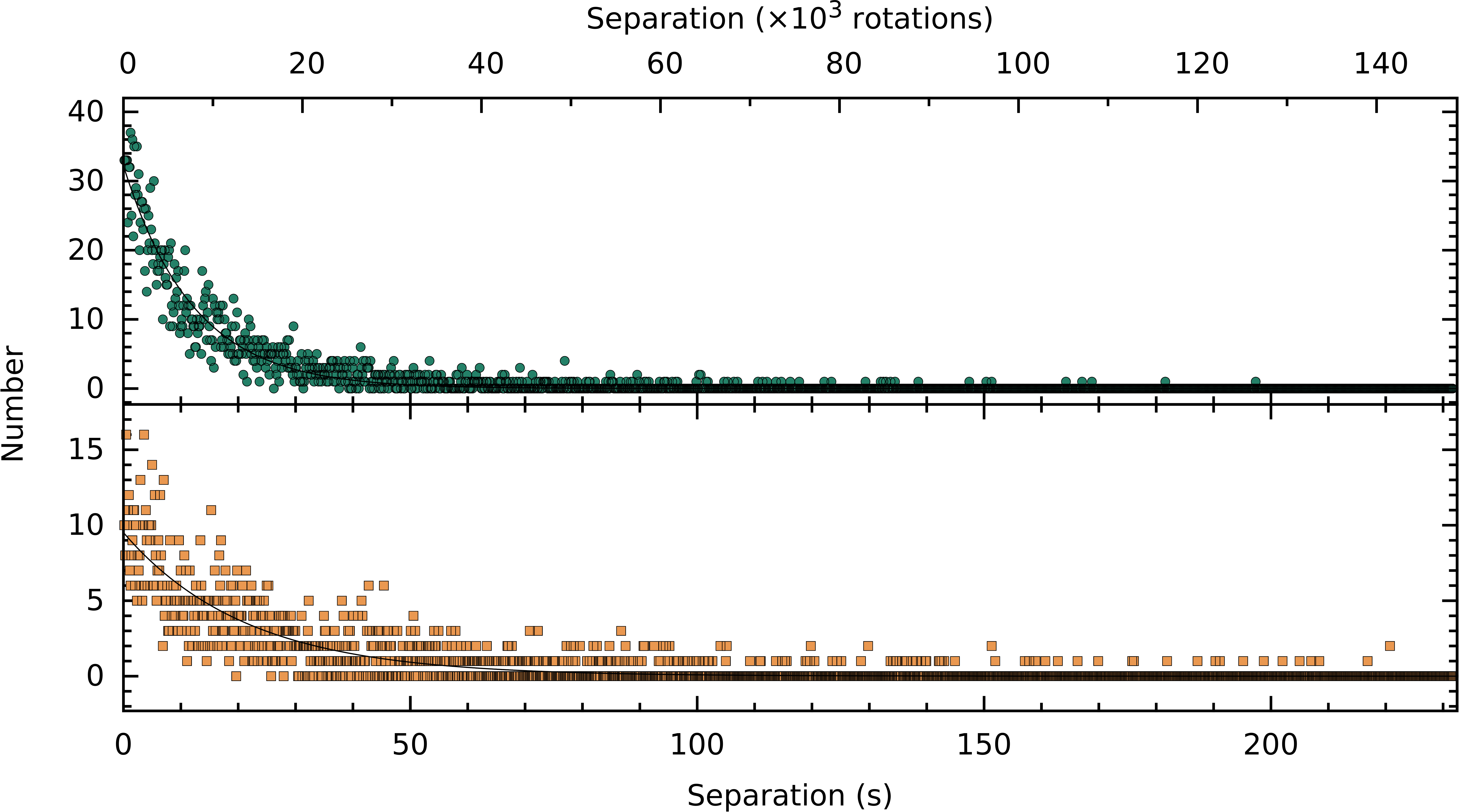}
	\centering
	\caption{Distribution of time intervals between successive MGPs (above, green circles) and IGPs (below, orange squares) for our entire data set, and the best-fit exponential functions to these distributions (solid lines). The distributions are well-described by an exponential decay, where the decay constant implies a mean GP rate of one MGP every $7813\pm61$ rotations ($12.2\pm0.1$\,s) and one IGP every $13699\pm188$ rotations ($21.4\pm0.3$\,s).}
	\label{fig:successiveGPs}
\end{figure*}

While the pulse shapes of the MGPs and IGPs do not appear to be correlated, there does appear to be a correlation on short time scales between GPs from the same emission region (Figure \ref{fig:gp_scint_structure_MP} and Figure \ref{fig:gp_scint_structure_IP}). The time scale of the GP microstructure was investigated, following the approach used by \protect\cite{kjv02} for regular emission, where periodicities in the sub-pulse emission are identified as peaks in the autocorrelation of individual GPs. Through this, the mean time scale of the separation between GP micropulses was found to be $249\pm197$\,ns, which follows the trend presented in \protect\cite{kjv02}, where the microstructure time scale is related to the pulsar spin period and is described by a power law (Figure \ref{fig:microscale}). The continuation of the scaling to a short-period pulsar suggests that GP emission could be composed of rotating mini-beams, which are likely the fundamental entities of emission, similar to the nano-shots observed in GP emission from the Crab Pulsar (\protect\citealp{hkw+03}, \protect\citealp{he07}).
As the PSR\,B1937+21 GP beam width is very much consistent with the scaling with period, we interpret this as evidence for beaming as a natural explanation for the observed short-duration of GPs.

\subsection{Giant Pulse Emission Rates} \label{emissionratessection}
Our large sample of GPs allows us to calculate the GP emission rate statistics. The statistics have applications in studies of the emission physics of GPs, and information about the typical wait time between GPs is useful in scintillometry, where GPs occurring within a time less than the decorrelation time scale are required to find the interstellar medium transfer function (e.g. \protect\citealp{mvp+17}).
Our observed emission rate from a large sample, and spanning a large number of widely-separated observations confirms that GPs from PSR\,B1937+21 occur very rarely. If we assume that the rates we calculate above are representative of the true emission rate, and that GPs are independent events, the probability of an IGP being emitted in a given rotation is ${\text{Pr}(\text{IGP})=4.7\times10^{-5}}$, while an MGP has $\text{Pr}(\text{MGP})=8.9\times10^{-5}$, and a GP from either emission region has $\text{Pr}(\text{GP})=1.4\times10^{-4}$. We did not observe an MGP and IGP in the same rotation, although this is not surprising as the probability of observing a so-called `double giant pulse' in our data set is only $\sim21\%$, using the values above. For a $95\%$ probability of observing a double giant pulse, $1.4\times10^{8}$ rotations would be required, approximately 4.5 times more than our data set.

\begin{figure*}
	\includegraphics[scale=0.26]{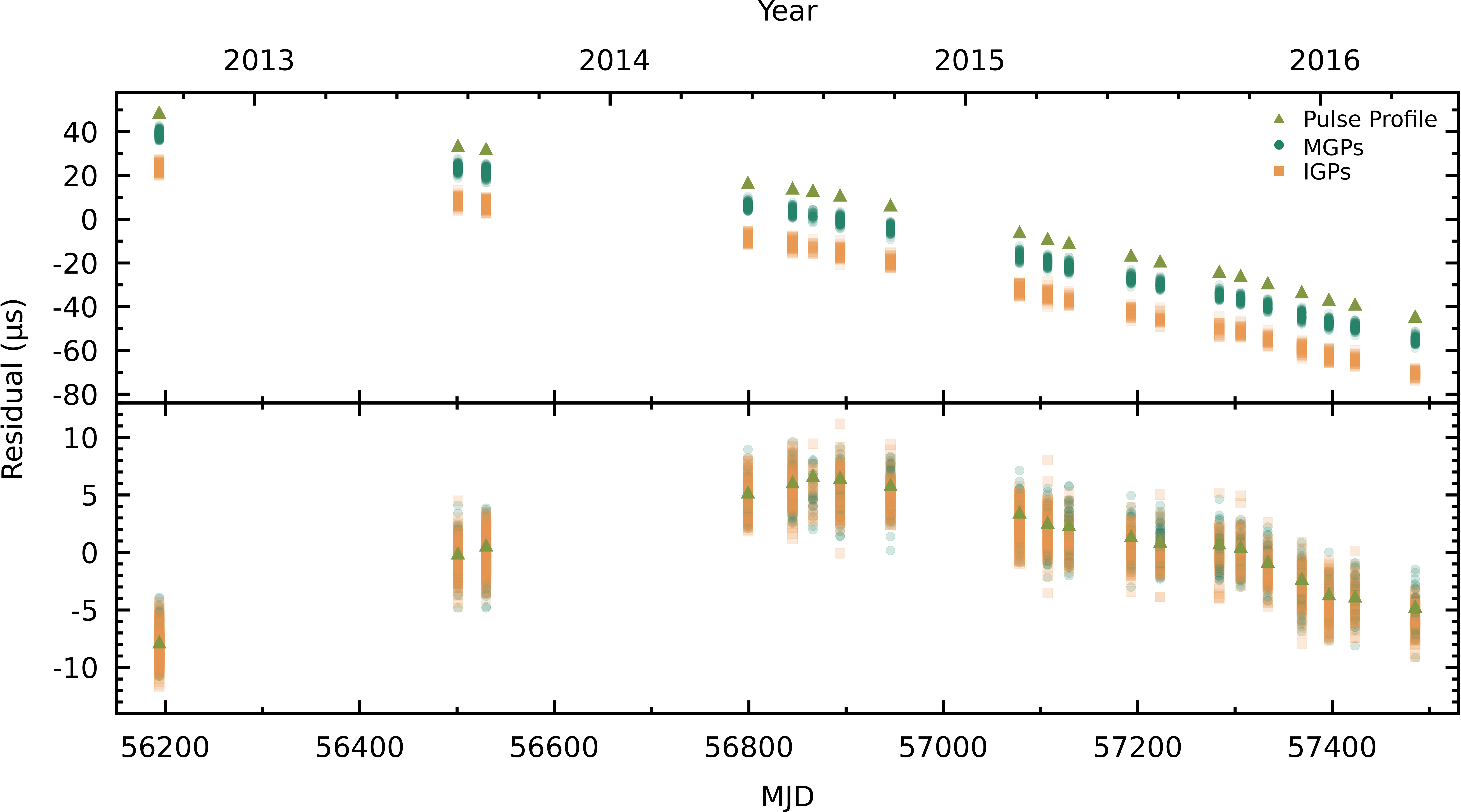}
	\centering
	\caption[GP timing residuals]{Timing residuals for the average pulse profile (yellow triangles), and GPs (green circles: MGP, orange squares: IGP). \textit{Above:} Timing residuals using the EPTA data release 1.0 timing model \protect\citep{dcl+16}, with the MGP and IGP residuals offset from the average pulse residuals by $-$\SI{10}{\micro\second} and $-$\SI{25}{\micro\second} respectively for clarity. The weighted RMS of the residuals, using the EPTA model, is \SI{28.0}{\micro\second}, \SI{25.7}{\micro\second}, and \SI{28.8}{\micro\second} for the MGP, IGP, and average pulse profile data sets respectively. The weighted RMS of the average-profile residuals is consistent with that of 4-yr subsets of the \protect\cite{dcl+16} data set. As expected, the spread of GP residuals is consistent with the width of the folded GP profiles (Figure \ref{fig:averageprofile}). \textit{Below:} Timing residuals aligned using constant phase offsets between the three data sets, with a linear term removed to aid clarity.}
	\label{fig:GPtiming}
\end{figure*}

The probability of $n$ GPs occurring within $m$ rotations in a data set of $N_{r}$ rotations can be written as
\\
\\
$\text{Pr}(n,m,N_{r})=$ 
\begin{equation} \label{GPocurrenceprobability}
 1-\left \{1-\left(\text{Pr}(\text{GP})^{n}[1-\text{Pr}(\text{GP})]^{m-n}\frac{m!}{n!(m-n)!}\right)\right \}^{N_{r}}.
\end{equation}
Equation \ref{GPocurrenceprobability} allows us to estimate the required length of a data set before GPs in consecutive rotations will be observed. Our data set contains no GPs in adjacent rotations, and according to Equation \ref{GPocurrenceprobability}, the probability of two GPs occurring within two rotations in $N_{r}=3.1\times10^{7}$ is $\text{Pr}(2,2,3.1\times10^{7})=46\%$, and again it is therefore not surprising that no GPs in consecutive rotations were observed. For a $95\%$ probability of observing two GPs occurring in consecutive rotations, a data set containing $1.5\times10^{8}$ rotations would be required, equivalent to $\sim66$\,hrs of observations, or a factor of 4.8 more than our data set, which is similar to the expected data set size required for a simultaneous MGP and IGP to be observed. The shortest separation in our sample is three rotations, between two MGPs observed on 25$^{\text{th}}$ August 2013 (MJD\,56529). These two closely-separated MGPs occur at different pulse phases (separated by $\sim$0.0015), but have some similarities: their pulse energies are 1.7\,Jy\,$\SI{}{\micro\second}$ and 1.8\,Jy\,$\SI{}{\micro\second}$, and their measured scattering time scales are consistent with zero. The frequency-dependent structure is also very similar for the two GPs, with the brightest emission occurring in the 1356\,MHz sub-band.

It has been proposed that GPs result from a self-organised criticality process \protect\citep{btw88}, where the underlying emission mechanism reaches some critical point, causing an `avalanche process' which results in GP emission (e.g. \protect\citealp{cai04}, \protect\citealp{mpw07}). Assuming this to be true, and that GP emissions from both components are independent events, the distribution of separations is expected to follow a Poisson probability distribution \protect\citep{lcu+95}, which gives rise to an exponentially-decreasing time between pulses from the most likely time separation.
We confirm that the expected distribution is followed for GPs from PSR\,B1937+21, by plotting the IGP time separations and MGP time separations separately, binned in intervals of 100 rotations, and fitting an exponential function ${N(t)=N_{0}\exp(-\lambda t)}$ to the distribution (Figure \ref{fig:successiveGPs}). This provides a good fit, with decay constants $\lambda_{\text{MGP}}=1.28\pm0.01\times10^{-4}$ and ${\lambda_{\text{IGP}}=7.3\pm0.1\times10^{-5}}$. From the exponential function, the mean separation between successive GPs can be estimated from the inverse of the decay constant as $\lambda_{\text{MGP}}^{-1}=7813\pm61$ rotations ($12.2\pm0.1$\,s) and $\lambda_{\text{IGP}}^{-1}=13699\pm188$ rotations ($21.4\pm0.3$\,s).

\subsection{Fast Radio Bursts and Undiscovered Pulsars}
It has been proposed that fast radio bursts (FRBs) could be explained by the very brightest GPs from pulsars located at cosmological distances (e.g. \protect\citealp{cw16}).
The lowest-inferred luminosity distance estimate of an FRB listed in the Fast Radio Burst Catalogue\footnote{\url{http://www.frbcat.org/}} \protect\citep{pbj+16} is $\sim0.54$\,Gpc (FRB\,170827; \protect\citealp{ffj+17}). 
The occurrence rate for FRBs at 1400\,MHz with pulse energies of 130--1500\,Jy\,$\SI{}{\micro\second}$ has been estimated to be {$7^{+5}_{-3}\times10^{3}$ events sky$^{-1}$\,day$^{-1}$} \protect\citep{cpk+16}. 
From the power law index we measure for the high-energy GPs, and assuming a distance to PSR\,B1937+21 of 3.27\,kpc \protect\citep{dcl+16}, the wait time for a 130\,Jy\,$\SI{}{\micro\second}$ GP from a PSR\,B1937+21-like pulsar at 0.5\,Gpc is $6.2\times10^{23}$\,hours, or $7.1\times10^{19}$\,years. For the lower limit of the \protect\cite{cpk+16} rate to be explained by a population of GP-emitting PSR\,B1937+21-like pulsars, a population of $\sim10^{26}$ would be required, which appears to rule out PSR\,B1937+21-like pulsars as the cause of the observed occurrence rate of FRBs. For this rate to be achieved, additional flattening of the pulse energy distribution at high pulse energies, as is the case with so-called `super-giant pulses' observed in the Crab Pulsar \protect\citep{cbh+04}, would be required.

An undiscovered population of pulsars is thought to exist in the Galactic Centre (GC; distance $\sim8.3$\,kpc, \protect\citealp{gef+09}, \citealp{wcc+12}) of which it is estimated $\sim10^{4}$ are detectable from the Earth (i.e. those whose beams cross the line of sight from the Earth, \protect\citealp{rla17}). Detection of this population may be possible via single-pulse searches for the GP-emitting pulsars. Assuming the single-pulse detection limit of 130\,Jy\,$\SI{}{\micro\second}$ from \protect\cite{cpk+16} used above, the wait time for a GP from a PSR\,B1937+21-like pulsar at the distance of the GC is 5376\,hours. If the fraction of GP-emitters in the GC is the same as that of the known pulsar population ($\sim0.005$, Table \ref{tab:GPemitters}), and their pulse energies all follow a PSR\,B1937+21-like power law, the wait time for a detectable GP from the GC is 108\,hours. However, in-depth searches for GPs from known pulsars have not been made, and in fact the true fraction of GP emitters is likely to be much higher.

\subsection{Timing of Giant Pulses} \label{timingsection}
GPs are characterised by their extremely short duration compared to the width of the average pulse profile (typically a factor of 100 narrower in the case of PSR\,B1937+21, with a GP duty cycle of $\delta\sim0.001$). As small duty cycles are related to TOA error by $\sigma_{\text{TOA}}\propto \delta^{3/2}$, GPs offer the opportunity to measure TOAs to much higher precision than is possible with the average pulse profile. This also has the advantage of providing many timing measurements in a given observation, allowing small-scale timing effects to be better isolated. 
The prospect of improving the PSR\,B1937+21 timing precision is particularly interesting, as it is one of two MSPs, along with PSR\,B1821$-$24A (PSR\,J1824$-$2452A) which is also a GP emitter \protect\citep{rj01}, for which timing noise is easily measureable (e.g. \protect\citealp{cd85}). This limits the timing stability on time scales greater than a few weeks \protect\citep{ktr94}.
As mentioned earlier, the non-GP single-pulse emission from PSR\,B1937+21 shows no evidence for pulse-shape variations \protect\citep{jap01}, which implies that irregularities in the spin of the pulsar due to e.g. intrinsic spin instabilities or the presence of an asteroid belt \citep{scm+13} are the cause of the observed timing noise. In contrast, other MSPs have been observed to display significant jitter noise through pulse-to-pulse variations (e.g. \protect\citealp{sod+14}, \citealp{lkl+12}), sub-pulse drifting (e.g. \protect\citealp{es03}, \citealp{lbj+16}), or variations in flux density (e.g. \protect\citealp{wmj07}).

To study the timing properties of GPs from PSR\,B1937+21, TOAs were generated from the polarisation- and frequency-averaged profiles of each of the GPs, by using standard timing techniques (e.g. \citealp{tay93b}). To account for the variable pulse shapes exhibited by the GPs (Figure \ref{fig:gp_scint_structure_MP} and Figure \ref{fig:gp_scint_structure_IP}), a noise-free template was made for the MGPs and IGPs separately, based on the highest-S/N in both of these categories. This template was found to describe the GPs well, with a typical TOA error of 3\,ns and a maximum TOA error of 63\,ns. 
TOAs were also generated using the average pulse profile of each of the observations, using a noise-free artificial reference template based on the highest-S/N observation of PSR\,B1937+21 with LEAP. The MGP, IGP, and average pulse profile timing residuals were analysed using an ephemeris derived from that presented in \protect\cite{dcl+16}, and a constant offset was fit between all three data sets. 

Residuals from the average pulse profile TOAs were found to have a weighted RMS consistent with that of 4-yr subsets of the \protect\cite{dcl+16} data set. The TOAs from each observation did not necessarily use the same centre frequency (Table \ref{tab:b1937observations}), and were corrected for DM variations by fixing the value to that found from the multi-frequency timing from the Lovell Telescope and the 42-ft telescope (see Section \ref{GP_pipeline}). It was found that the DM values used to minimise the GP widths added significant structure to the residuals, mirroring the variation of these values seen in Figure \ref{fig:B1937multifreqdm}, which indicates that the DM values used to optimise the GP search do not accurately describe the delay seen in the timing data. This may be due to some additional or variable dispersion of GP by the pulsar magnetosphere, as seen in the Crab Pulsar \protect\citep{he07}, or due to frequency-evolution of the GP pulse shapes (Figure \ref{fig:gp_scint_structure_MP} and Figure \ref{fig:gp_scint_structure_IP}).

We find that although the TOA errors for the individual GPs used in our timing data set are much lower than those of the average pulse profiles (with the lowest average pulse profile TOA error being 7\,ns), the timing precision is not significantly different when using the GPs, with the resulting weighted RMS of the residuals being \SI{28.0}{\micro\second}, \SI{25.7}{\micro\second}, and \SI{28.8}{\micro\second} for the MGP, IGP, and average pulse profile TOAs respectively, and mean residuals weighted by number of TOAs in an observation of 13.6\,ns for the MGPs and 26.2\,ns for the IGPs (Figure \ref{fig:GPtiming}). Although using GPs to time PSR\,B1937+21 does not offer a significant improvement in timing precision, more rotationally-stable GP-emitting pulsars (e.g. PSR\,J0218+4232) may still benefit from this approach. The spread of the GP residuals is as expected from the pulse width of the average GP profile, where the rate and amplitude of individual GPs leads to the average GP shape (Figure \ref{fig:averageprofile}). Using average GP profiles from each observation, we measure the mean widths of the MGP and IGP distributions to be $4\pm\SI{1}{\micro\second}$ and $4\pm\SI{2}{\micro\second}$ respectively, which are consistent with the values reported by \protect\cite{kt00}.

\section{Conclusions} \label{conclusions_section}
We have searched for GPs in baseband data from 13.6\,hrs of observations of PSR\,B1937+21 with the LEAP telescope, finding a total of 4265 GPs, which is the largest-ever sample of GPs for this pulsar. We do not observe GPs in consecutive rotations, but calculate that this is not surprising, given the size of our data set. By using a modified version of the radiometer equation, we have estimated the pulse energy for each of the GPs, which we have used to comment on the pulse energy distributions and emission rates. We have measured the scattering influence on the pulse shapes of individual GPs and found no correlation between mean scattering time scale and DM. We find that individual GPs generally have higher fractional polarisations than that of the average pulse profile, and find that there is no phase-dependence of the polarised emission from within the GP emission region. We use the GPs to time PSR\,B1937+21, and find that although the achievable TOA precision is much higher for GPs than for the average pulse profile, the weighted RMS of the GP timing residuals does not offer a significant improvement over those of the average pulse profile.

We note that the GP emission rate and pulse energy distribution power law indices vary considerably between observations, as does the structure of individual GPs. These properties are in contrast to the high stability observed in the regular pulse emission \protect\citep{jap01}, and are in agreement with the significant measurement of pulse modulation for the GP pulse stacks (Figure \ref{fig:GPmodulationindex}). The difference in these properties between the regular emission and the GP emission, and the fact that we see regular emission and GP emission can occur simultaneously (Figure \ref{fig:averageprofile}) supports the hypothesis that GP emission arises from a separate process to that of the regular emission.

\section*{Acknowledgements}
We thank Patrick Weltevrede, Bhaswati Bhattacharyya, Cristina Ilie, Axel Jessner, Robert Wharton, Joris Verbiest, and Aris Noutsos for valuable discussions. This work is supported by the ERC Advanced Grant ``LEAP", Grant Agreement Number 227947 (PI M.\,Kramer). The European Pulsar Timing Array (EPTA) is a collaboration between European Institutes, namely ASTRON (NL), INAF/Osservatorio di Cagliari (IT), the Max-Planck-Institut f{\"u}r Radioastronomie (GER), Nan{\c c}ay/Paris Observatory (FRA), The University of Manchester (UK), The University of Birmingham (UK), The University of Cambridge (UK), and The University of Bielefeld (GER), with an aim to provide high-precision pulsar timing to work towards the direct detection of low-frequency gravitational waves.  Access to the Lovell Telescope is supported through an STFC consolidated grant. The Effelsberg 100-m telescope is operated by the Max-Planck-Institut f{\"u}r Radioastronomie. The Westerbork Synthesis Radio Telescope is operated by the Netherlands Foundation for Radio Astronomy, ASTRON, with support from NWO. The Nan{\c c}ay Radio Observatory is operated by the Paris Observatory, associated with the French Centre National de la Recherche Scientifique. KJL gratefully acknowledge support from National Basic Research Program of China, 973 Program, 2015CB857101 and NSFC 11373011. KL and RK also gratefully acknowledge support from ERC Synergy Grant ``BlackHoleCam", Grant Agreement Number 610058  (PIs: H.\,Falcke, M.\,Kramer, L.\,Rezzolla).

\bibliographystyle{mnras}
\bibliography{psrrefs}{}
\bsp	

\label{lastpage}
\end{document}